\documentclass[journal]{IEEEtran}

\usepackage{graphicx}
\usepackage{cite}
\usepackage{amsmath,amssymb,amsfonts}
\usepackage{multirow}
\usepackage{hyperref}
\usepackage{subcaption}
\usepackage{placeins}
\usepackage{color}
\usepackage{soul}
\usepackage{textcomp}
\usepackage{array}
\usepackage{stfloats}
\usepackage{makecell}

\renewcommand\hl[1]{#1}  
\begin{document}
\title{Collection and Validation of Psychophysiological Data from Professional and Amateur Players: \\a Multimodal eSports Dataset }
%
%
%
\author{
	\IEEEauthorblockN{
		\textsuperscript{1}Anton Smerdov,
		\textsuperscript{2}Bo Zhou,
		\textsuperscript{2}Paul Lukowicz,
		\textsuperscript{1}Andrey Somov
		}
	
	\IEEEauthorblockA{
	\textsuperscript{1}Skolkovo Institute of Science and Technology, CDISE, Moscow, Russia}
	
	
	\IEEEauthorblockA{
	\textsuperscript{2}German Research Centre for Artificial Intelligence, Kaiserslautern, Germany}
	}

\markboth{IEEE Transactions on Games,~Vol.~xx, No.~y, Month~20xx}%
{Smerdov \MakeLowercase{\textit{et al.}}: Mutlimodal eSports Dataset}

\maketitle

\begin{abstract}

Proper training and analytics in eSports require accurately collected and annotated data.
Most eSports research focuses exclusively on in-game data analysis,
and there is a lack of prior work involving eSports athletes' psychophysiological data.
In this paper, we present a dataset collected from professional and amateur teams in 22 matches in League of Legends video game \hl{with more than 40 hours of recordings}.
Recorded data include the players' physiological activity, e.g. movements, pulse, saccades,
obtained from various sensors, self-reported after-match survey, and in-game data.
An important feature of the dataset is simultaneous
data collection
from five players, which facilitates the analysis of sensor data on a team level. Upon the collection of dataset we carried out its validation. In particular, we demonstrate that stress and concentration levels for professional players are less correlated, meaning more independent playstyle. Also, we show that the absence of team communication does not affect the  professional players as much as amateur ones.
To investigate other possible use cases of the dataset,
we have
trained classical machine learning algorithms for skill prediction and player re-identification using 3-minute sessions of sensor data. Best models achieved 0.856 and 0.521 (0.10 for a chance level) accuracy scores on a validation set for skill prediction and player re-id problems, respectively. The dataset is available at \url{https://github.com/smerdov/eSports_Sensors_Dataset}.

\end{abstract}


\IEEEpeerreviewmaketitle

\section{Introduction}

Competitive video gaming, or eSports, has gained tremendous popularity within the last several years.
eSports has evolved into a mature industry
with well-funded tournaments, professional athletes, and a vast fan community.
The strengthened competition requires professional eSports teams to explore new
methods
for training and analytics.
In its turn, it drives the demand for eSports research.



A common and natural source of data for eSports research is in-game data.
The majority of prior research relies on information obtained from the game logs, e.g. heroes drafted, abilities learned, players' positioning~\cite{dota_live_prediction_, dota_draft_prediction_1, lol_live_prediction_1}.
This information may be helpful in
evaluating the players' performance,
predicting the outcome of the match, 
understanding players' behavior,
and other analytics.
However, if only in-game data are used,
much information from the real world is omitted. Sensor data obtained from the physical domain can supplement in-game logs collected in the digital domain~\cite{eye_tracking_patterns_2, esports_skill_prediction, esports_gsr}.
Similarly, predictive models trained on sensor data can supplement regular models trained using in-game data and provide additional evidence for analytics~\cite{gaze_reactions, smart_chair_iop}.
Data obtained from sensors can represent players' physical conditions, psychological state,
environmental situation,
and other factors
that may help in eSports analytics. 
Investigation of the connection between the 
physiological data
and player performance
may help provide personalized feedback for efficient training.

The major disadvantage of using only in-game data is the poor robustness of predictive models. These models capture dependencies between the  in-game parameters and some target values.
When rules or game mechanics change, previous dependencies
utilized to fit the model
might not persist.
Given the regularity of patches and updates in eSports games versions, a tight connection between the in-game parameters and model prediction can significantly limit model usage lifespan.
It makes the models based on real-world data collected from sensors practically feasible since they are not sensitive to game mechanics changes.

Although the idea to use sensor data in eSports research 
is not new, prior work on this matter is typically limited to one sensor used or one player recorded~\cite{esports_eeg_0, identification_by_mouse, esports_visual_fixations}.
That narrows the insights explored as the
multi-sensor dependencies are not captured, and person-to-person interactions are not registered.
Recording with multiple sensors might help understand which signal is more important in the eSports domain and improve the model performance
and robustness since the data from multiple sources are utilized.
Capturing sensor data from multiple players simultaneously can help
capture interactions between players
and investigate team dynamics, which is especially relevant because many eSports games are team games.

In this work, we present a dataset of sensor data collected from 10 participants playing League of Legends, 22 matches in total.
Participants were invited based on their skill level and organized in two teams of 5 players:
professional players with vast experience in the game and amateur players with basic experience in League of Legends.
The sensor data collected include information on hand/head/chair movements, heart rate, muscle activity,
gaze movement on the monitor,
galvanic skin response, electroencephalography (EEG), mouse and keyboard activity, facial skin temperature, and environmental data. Additionally, we provide
feedback from each player after each match with their feeling about each match. Every match is accompanied by a game log indicating key in-game events like kills/deaths/assists, players positioning, skills learned, buildings destroyed, etc. 
The dataset is structured as directories for each of 22 matches, and each match consists of directories for each of 5 players with many \texttt{.csv} files with sensor data presented.


We consider the extreme setting when only sensor data are available for prediction, and in-game data are used for labeling only. It allows us to train the model that is independent from in-game parameters
and check if sensor data are enough to properly understand the players' behavior.


Contributions of this research is twofold:

\begin{enumerate}
    \item An extensive dataset with sensor data collected from players with various skill levels.
    \item A team dynamics study in the eSports domain to show the validity of data collected.
\end{enumerate}

In terms of novelty, we carry out the data collection from \textit{professional} players (apart from the amateur ones) recording the data from five players at a time. The collected data are truly multi-modal including the physiological data, in-game data, and self-reported survey.

This paper is organized as follows:
in Section II we overview the state-of-the-art methods using sensor data for performing the data  analysis in eSports; in Section III we present a sensor network architecture for data collection; in Section IV we describe the methodology of data collection; in Section V we describe in details the data collected; in Section VI we provide concluding remarks.

\section{Related Work}

State-of-the-art work in eSports research can be divided into two categories:
(i) a wide group of studies exploring dependencies in in-game data,
and (ii) a small scope of work investigating the importance of psychophysiological signals in the eSports domain.
In this section we review methods utilizing sensor data in eSports
and discuss competing approaches which use data only from the digital domain.





\subsection{Psychophysiological Research in eSports}




Although psychophysiological research in eSports is still in its infancy, there is a number of prior studies utilizing one or a few sensors for data collection.

A popular sensor for eSports research is an eye tracker. Using the data obtained from an eye tracker, authors in \cite{esports_visual_fixations}
suggest that the crucial trait of professional players might be the duration and variability of visual fixations.
This idea is elaborated further in \cite{gaze_reactions} where authors propose to classify players into professional and amateurs based on their reaction time. They calculate the whole reaction time as a sum of three components: saccadic latency, the time between saccade and fixation, and the time for aiming and shooting.
In other studies relying on eye tracking for eSports \cite{eye_tracking_patterns_1}\cite{eye_tracking_patterns_2} the authors investigate the eye movement patterns able to distinguish the  professional players from amateurs. According to their results, professional players' eye-movement patterns are more diverse and swifter.

Another data source in eSports research is EEG. Minchev et al.\cite{esports_eeg_0} showed the decrease of EEG alpha rhythms power spectra frequencies for ‘lose’ game events and increase for the ‘win’ game events in a first-person shooter (FPS) game;
the opposite was shown for the theta rhythm. Another work on EEG in eSports \cite{esports_eeg_1} claims that the EEG signal for newbies has more variations than one for experienced players, although only two participants are compared.

Players' movements data can also help to investigate their behavior.
In \cite{smart_chair_iop} and \cite{smart_chair_wf_iot} authors showed that data about chair movements can help to separate high-skilled and low-skilled players in Counter-Strike: Global Offensive. They claim that professional players move on the chair less often but more intensively.

Game input pattern obtained from mouse or keyboard is also used for eSports research. In \cite{identification_by_mouse}, the authors showed that mouse input data might be useful for player re-identification. Work by Khromov et al.~\cite{esports_skill_prediction} investigated the connection between player's skill and data from mouse, keyboard, and eye tracker. Data obtained from the keyboard turned out to be the most important.
Work by Blom et al.\cite{esports_gsr} presented a dataset with mouse, keyboard, and galvanic skin response
for League of Legends players but did not show any dependencies in the data. 
In another work \cite{lol_streams_emotions} dedicated to League of Legends,
authors presented a dataset with stream videos annotated with streamer affect and game context.
They managed to predict the streamer's arousal and game context
using convolutional neural networks.


The above mentioned research demonstrate that sensor data can be used in a wide range of eSports problems, such as skill prediction, player re-identification, performance evaluation, etc.

\subsection{In-game Data in eSports}

Most of the prior eSports studies rely exclusively on the in-game data collection and further analysis.
The straightforward approach is to use in-game indicators such as kills, deaths, and gold to predict match outcome in Multiplayer Online Battle Arena (MOBA) \cite{dota_live_prediction_, dota_live_prediction_0, dota_live_prediction_1, lol_live_prediction_0, lol_live_prediction_1} or 
FPS genres \cite{fps_prediction_0}.
More advanced methods also try
to mine logic from a predictive model \cite{lol_logic_mining, lol_prediction_interpretation} for better interpretation, 
to extract high-level features (i.e., encounters statistics) from game replays \cite{encounter_prediction},
or
to calibrate the confidence of prediction \cite{confidence_prediction_lol}.

Another approach for match outcome prediction
is to use information about the hero drafts to predict a winner before the actual game starts \cite{dota_draft_prediction_0, dota_draft_prediction_1, dota_draft_prediction_2, lol_draft_prediction_0}. These methods can be easily transformed into draft recommendation systems with high practical use.

Genre-agnostic methods for eSports analytics usually rely on match history data and utilize only match results instead of ad-hoc features for a specific game.
In \cite{phys_momentum}, the authors leverage players' match history data to estimate players' current psychological state and use this estimation to predict the outcome of matches. 
Other history-based outcome prediction approaches include machine learning on hand-crafted features for the MOBA genre
\cite{history_prediction_1}
and modeling with rating systems for MOBA and shooter
genres \cite{history_prediction_2}.
Work by Dehpanah et al. \cite{ffa_rating_system} extends rating systems for free-for-all games (Battle Royale genre) using a large dataset of matches in PlayerUnknown’s Battlegrounds game.

Analytics in eSports in not limited to winner prediction but also extends to players' behavioral analysis.
Authors in \cite{esports_clusters} 
proposed to cluster players' in-game data in Massively multiplayer online role-playing games(MMORPG) and strategy games to interpret their behavior and to compose the optimal team lineup.
Research in \cite{gao} addresses the problem of player re-identification in the MOBA genre. The authors showed that classical machine learning algorithms can predict hero ID and one of three roles based on game data.
Pfau et al. \cite{dl_bot} emphasized the problem of players' Internet disconnection and
proposed to model the behavior of disconnected players with deep learning algorithms.
In \cite{toxic_behavior}, authors address the problem of toxic behaviors in the eSports games and argue that despite some effort from game developers, this problem persists in eSports community.
In
\cite{toxicity_detection}, authors trying to identify toxic behavior automatically with a toxicity detection algorithm based on the analysis of game chat logs.


\section{Sensor Network Architecture}

We used twelve types of sensors for data collection:

\begin{figure*}[!t]
    \centering
	\centerline{\includegraphics[width=\linewidth]{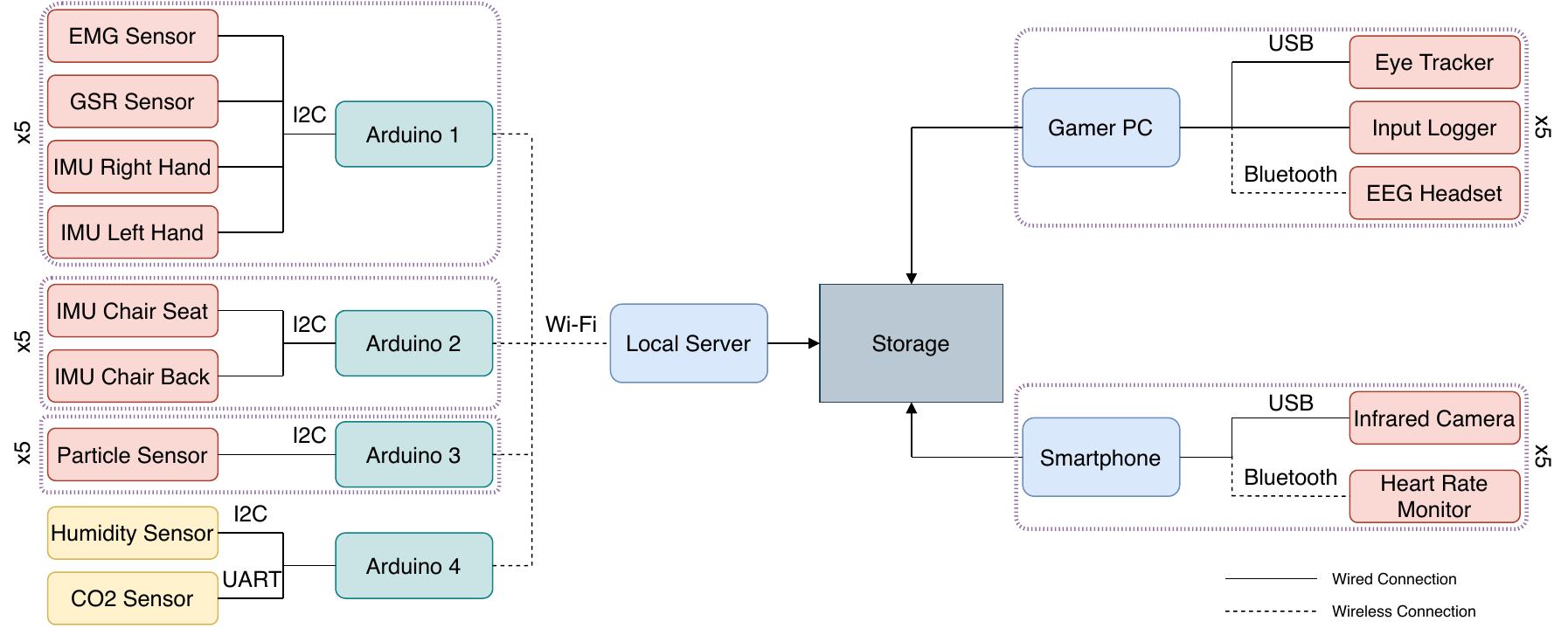}}
	\caption{Sensor network architecture. Solid lines are wired connections; dashed lines are wireless connections.}
	\label{sens_arch}
\end{figure*}

\begin{figure}[!b]
    \begin{subfigure}[!tp]{\linewidth}
	\centerline{\includegraphics[width=1.00\linewidth]{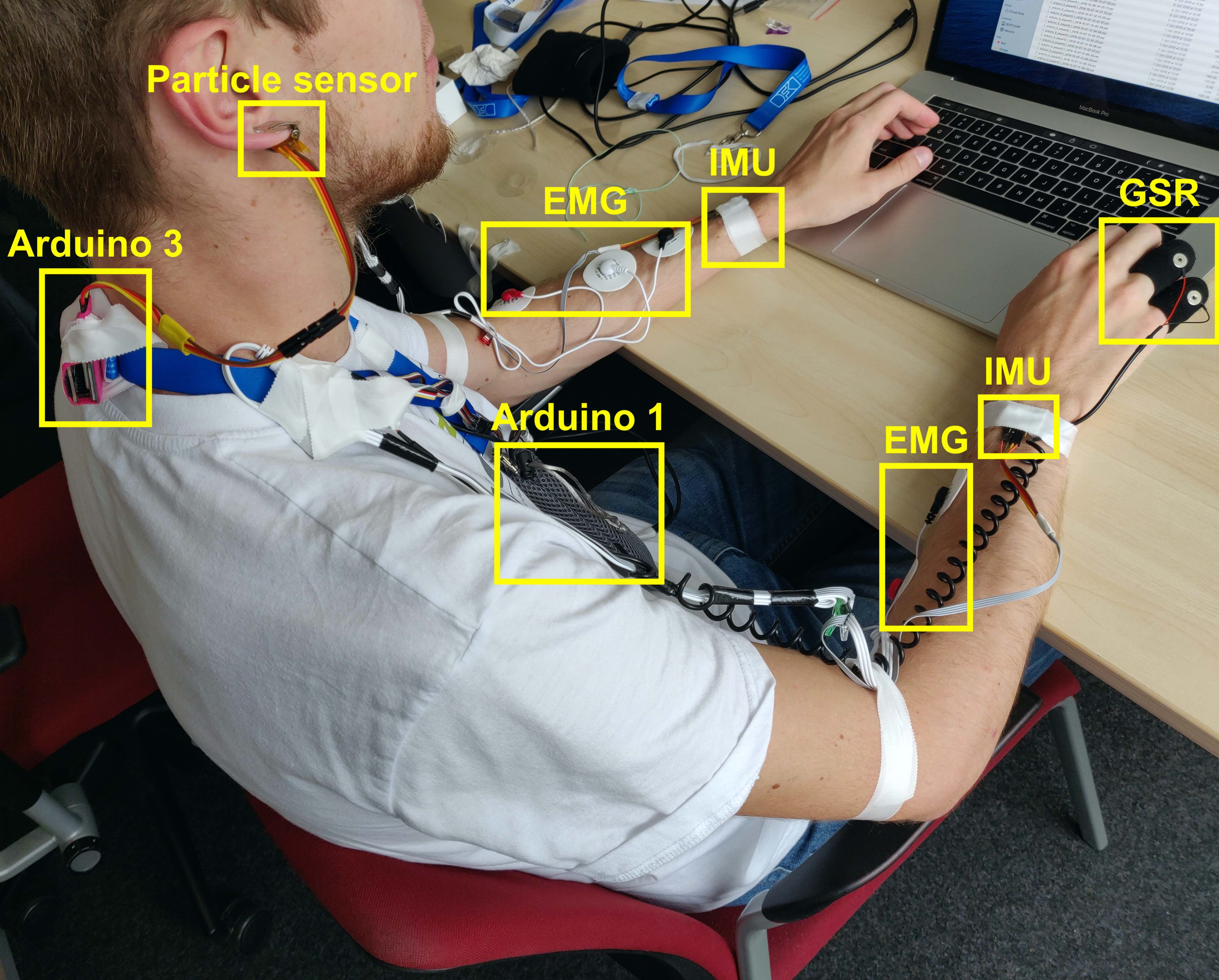}}
	\caption{Part of the sensing system: wearable sensors.}
	\label{setup_0}
	\end{subfigure}
	\begin{subfigure}[!tp]{0.365\linewidth}
	\centerline{\includegraphics[height=4.25cm]{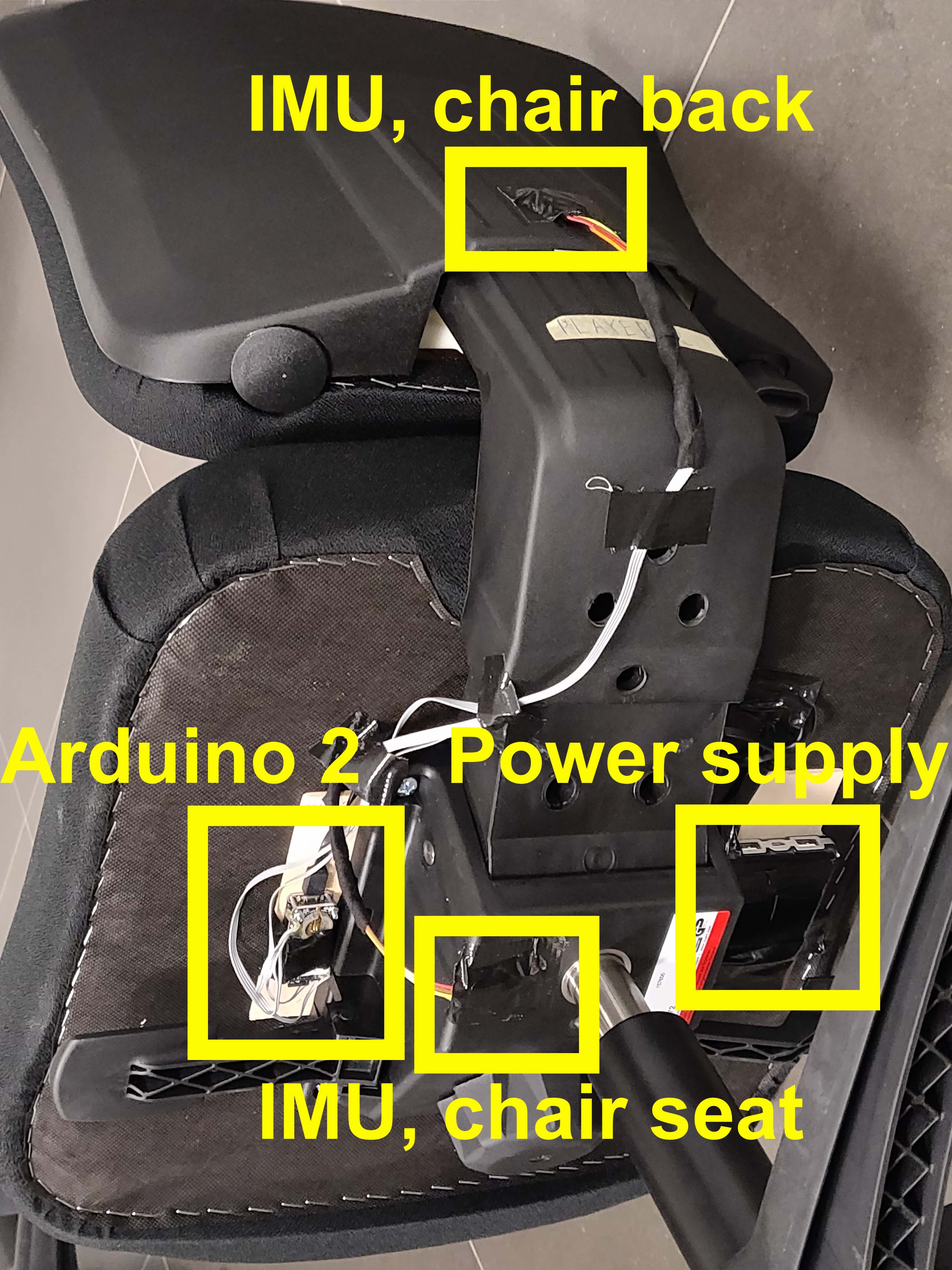}}
	\caption{IMU sensors integrated into a chair.}
	\label{chair_0}
	\end{subfigure}
	\label{overall_setup}
	\begin{subfigure}[!tp]{0.635\linewidth}
	\centerline{\includegraphics[height=4.25cm]{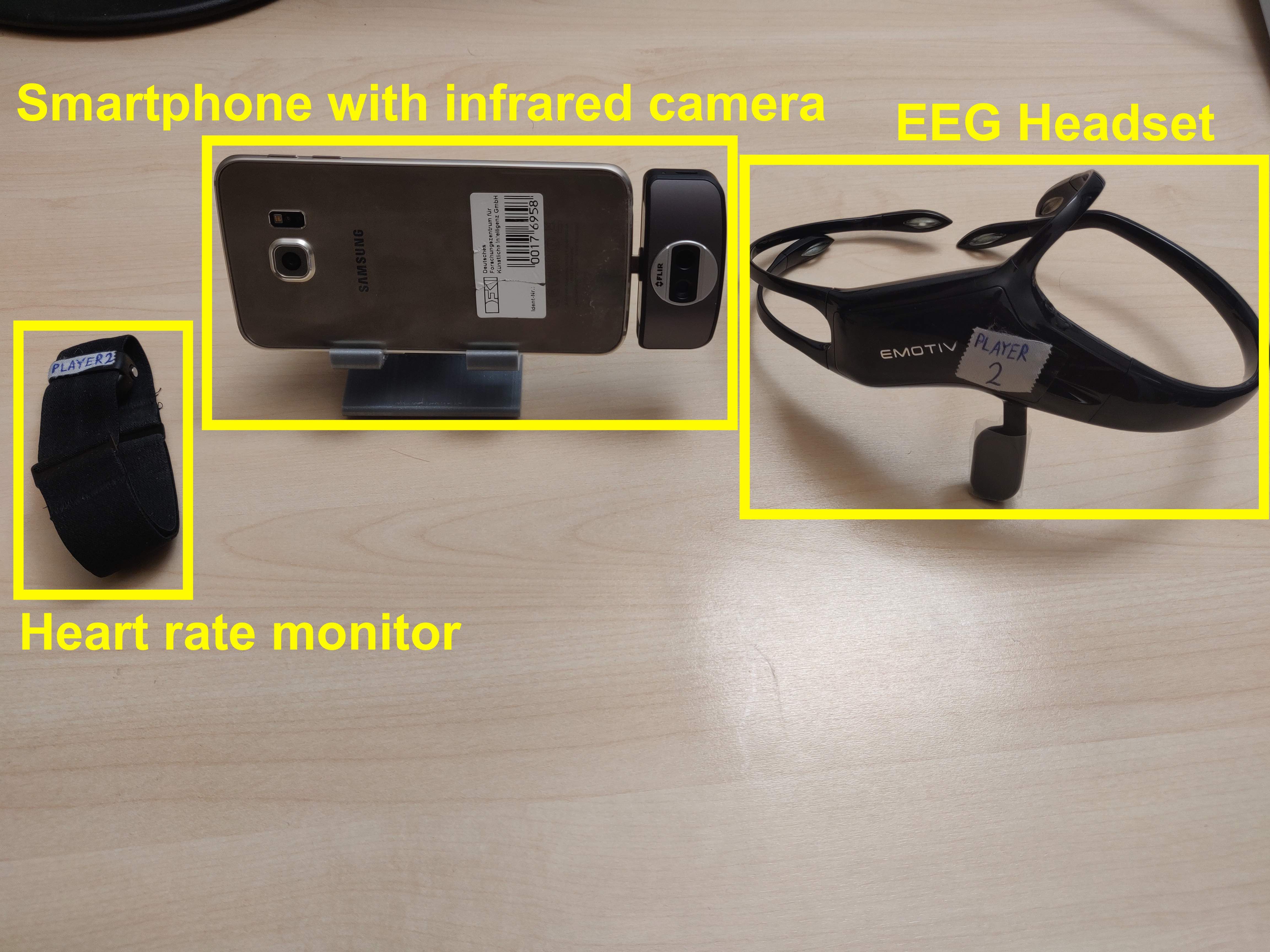}}
	\caption{Heart rate monitor, infrared camera, and EEG headset utilized.}
	\label{sensors_1}
	\end{subfigure}
	\caption{Experimental testbed.}
	\label{overall_setup}
\end{figure}

\begin{enumerate}
    \item Electromyography (EMG) sensor \textit{Grove EMG Sensor v1.1}\footnote{\url{https://wiki.seeedstudio.com/Grove-EMG_Detector/}}.
    This sensor measures the intensity of muscle activity
    and outputs it as voltage (analog interface). It uses ground, plus, and minus electrodes located on a forearm muscle to measure the voltage. The voltage can be interpreted as the intensity of muscle constraint; thus, the intensity of keyboard or mouse strokes or movement. EMG signal is related to physical tension affecting the player’s current state \cite{emg_state}.
    
    \item Galvanic skin response (GSR) sensor \textit{Grove GSR Sensor v1.2}\footnote{\url{https://wiki.seeedstudio.com/Grove-GSR_Sensor/}}. This sensor outputs voltage (analog interface), which allows us to infer skin resistance for the participant. Skin resistance is lower in the arousal state due to more sweat, so these data can be used as a stress indicator \cite{gsr_arousal}.
    Participants placed two electrodes of this sensor on two fingers on the right hand.
    
    \item Inertial measurement unit (IMU) \textit{Bosch BNO055}\footnote{\url{https://www.bosch-sensortec.com/products/smart-sensors/bno055.html}} located on the wrists on both hands. The sensor recorded linear acceleration, gravity, angular velocity, Euler angles, and quaternions.
    
    \item IMU's under the chair seat and on the chair back. Precise locations are shown in Figure~\ref{chair_0}. Behavior on a chair is connected with the player skill \cite{smart_chair_iop}\cite{smart_chair_wf_iot}. Sensors' models and data recorded are the same as from the IMU's located on hands.
    
    \item Environmental atmosphere sensor \textit{Bosch BME280}\footnote{\url{https://www.bosch-sensortec.com/products/environmental-sensors/humidity-sensors-bme280/}} located in the gaming room. It measures relative humidity and environmental temperature. These data affect players performance: high level of relative humidity results in the decrease of neurobehavioral performance \cite{humidity_performance}; high environmental temperature may affect human performance \cite{env_temperature_performance}.
    
    \item CO$_{2}$ sensor \textit{MH-Z19B}\footnote{\url{https://www.winsen-sensor.com/d/files/infrared-gas-sensor/mh-z19b-co2-ver1_0.pdf}}. This sensor measures carbon dioxide concentration in ppm. High CO$_2$ level may deplete cognitive abilities \cite{co2_cognitive}.
    
    \item Eye tracker Tobii 4C located under the monitor.
    Data from eye tracker indicate which information on the game screen players check more often, so the effectiveness of their game decisions \cite{esports_visual_fixations}. To standardize the information obtained with eye trackers, all players ran games in 1920x1080 resolution. Before each experimental day, players calibrated the eye trackers.
    
    \item Electroencephalography (EEG) headset \textit{Emotiv Insight}\footnote{\url{https://www.emotiv.com/insight/}}. EEG data have been shown to be connected with arousal \cite{eeg_arousal}, so the player performance.
    
    \item Input (keyboard and mouse) logger as a python script running on the gamer's PC. These data are indirect indicators of the hand movement activity and the player skill \cite{mouse_skill}.
    
    \item Infrared camera \textit{Flir One}\footnote{\url{https://www.flir.eu/flir-one/}} directed to players' faces for facial skin temperature data collection.
    
    \item Heart rate monitor \textit{Polar OH1}\footnote{\url{https://www.polar.com/en/products/accessories/oh1-optical-heart-rate-sensor}} armband. High heart rate corresponds to mental stress and arousal \cite{heart_rate_performance}, which might affect player decisions' rationality.
    
    \item Pulse-oximeter sensor \textit{Maxim MAX30102}\footnote{\url{https://www.maximintegrated.com/en/design/reference-design-center/system-board/6300}} located on the right earlobe. Prior research in \cite{oxygen_saturation_performance} demonstrated the connection between oxygen saturation and cognitive performance.
\end{enumerate}

\begin{table}[!t]
\centering
\caption{Sampling rates and missed values for sensor data.}
\label{sampling_rates}
\begin{tabular}{cccc}
Sensor name & \thead{Sampling rate,\\ Hz} & \thead{\hl{Hours}\\ \hl{collected}} & \thead{Data\\ Missed} \\\hline\hline
EMG sensor & \multirow{3}{*}{36} & \multirow{3}{*}{41.6} & \multirow{3}{*}{0\%}\\
GSR sensor & & & \\
IMU right\&left hand &  & \\\hline
IMU chair seat\&back & 65 & 40.1 & 3.6\%\\\hline
Particle sensor & 16-25 & 41.6 & 0\%\\\hline
Environmental sensor  & \multirow{2}{*}{1} & \multirow{2}{*}{41.6} & \multirow{2}{*}{0\%}\\
CO$_2$ sensor &  & \\\hline
Eye tracker & 90 & 34.0 & 18.3\%\\\hline
Mouse\&Keyboard & - & 35.3 & 15.1\%\\\hline
Heart rate monitor & 1 & 35.7 & 14.2\%\\\hline
\hline
\end{tabular}
\end{table}

Figure~\ref{sens_arch} illustrates the sensor network architecture. The experimental testbed is presented in Figures~\ref{setup_0}~and~\ref{chair_0}.
We used three Arduino boards to collect data from each person.
EMG and GSR sensors were connected to analog ports on the Arduino 1. Two IMUs on both hands were also connected to Arduino 1 but via I2C protocol. Arduino 2 gathered data from two IMUs located on the chair. Arduino 3 collected the signal from the particle sensor via I2C protocol.
Arduino 4 was used to collect environmental data. It received data from the humidity sensor via I2C protocol and from the CO$_2$ sensor via UART. Arduino boards were powered with standard 5.0V 10000 mAh batteries. In total, we used 16 Arduino boards and 10 batteries.

We used a separate smartphone for each player to manage data collection from a heart rate monitor and an infrared camera.
Heart rate monitors communicated with devices
via
Bluetooth standard; infrared cameras were plugged into Micro-USB or USB-C ports in smartphones.
Heart rate monitors were managed by an Android app \textit{Polar Flow}. Infrared cameras were managed by a custom app written using FLIR ONE SDK\footnote{\url{https://developer.flir.com/mobile/flironesdk/}}.


Each EEG headset transferred data to a separate PC via Bluetooth standard. Data transmitted were logged into files using a custom python script.
Data from each eye tracker were logged to the corresponding gaming PC.
Mouse and keyboard activity were also recorded on each gaming PC by a custom python script based on the \texttt{pyinput}\footnote{\url{https://pypi.org/project/pynput/}} library.


Before each experiment, we synchronized all Arduino boards and PCs with an NTP server.
Considering the desynchronization of devices through the experiment's duration, the difference in time
did not exceed
500 ms.

\hl{In total, we collected 54.6 hours of recordings.}
Sampling rates for sensors, \hl{durations of data collected,} and proportions for data missed are presented in Table~\ref{sampling_rates}. 
\hl{Excluding data from the EEG headset and infrared camera, the duration of data collected is more than 34 hours, and the proportion of data missed does not exceed 18.3\%. Duration of the data collected for EEG is 12.2 hours (70.6\% missed), for the infrared camera is 26.6 hours (36.1\% missed). We present these data in the dataset but don't use it for the analysis. The limitation for EEG data collection was a Bluetooth interference in the room. The limitation for data collection from infrared cameras was a battery life of the devices, because they couldn't be charged when the infrared camera is plugged in.}

After each experiment, we aggregated the data collected from Arduino boards, smartphones, and PCs to one storage device.

\section{Methodology}\label{section_methodology}

\begin{figure}[!t]
	\centerline{\includegraphics[width=0.95\linewidth]{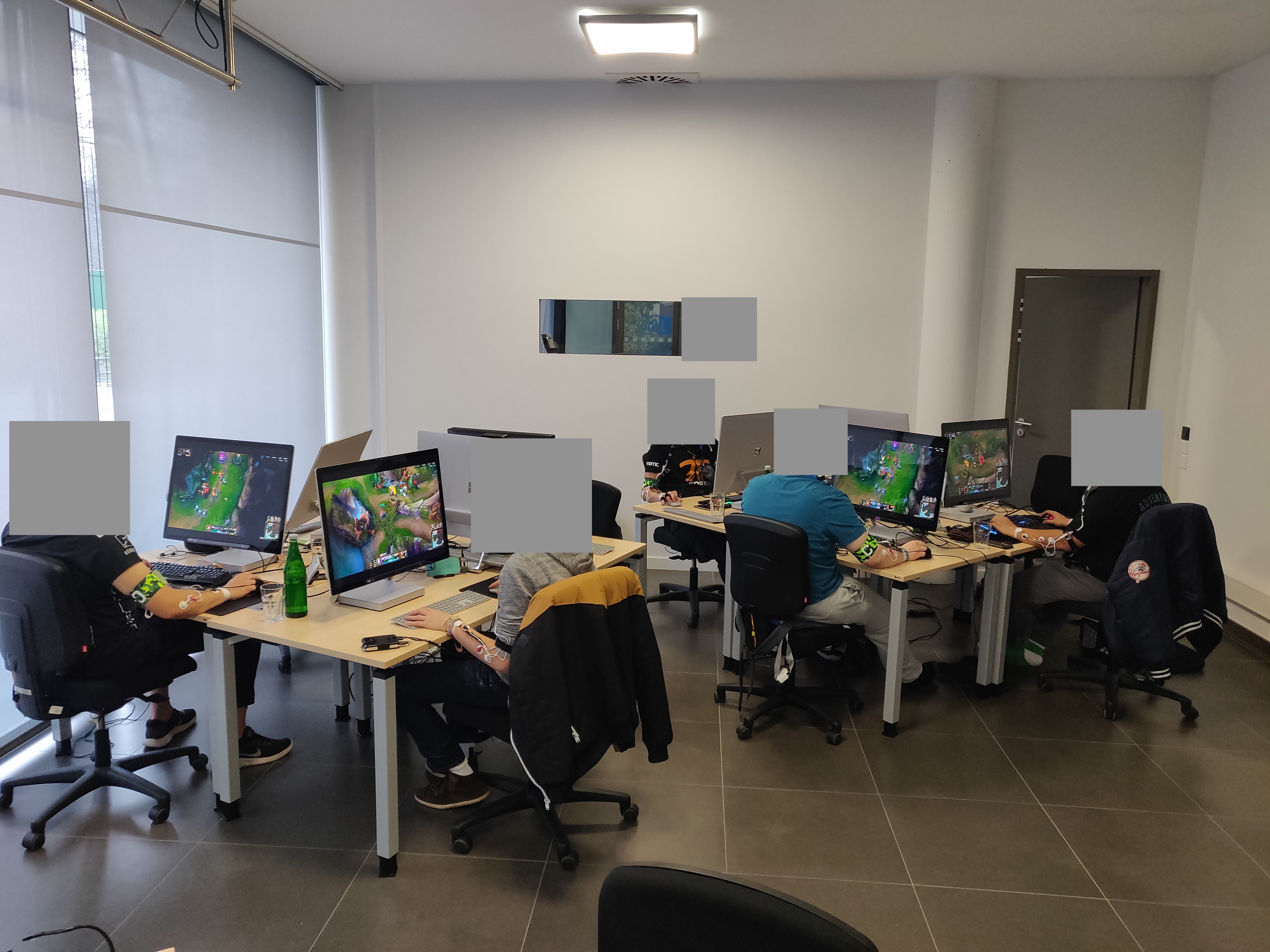}}
	\caption{Process of data collection.}
	\label{exp}
\end{figure}

We invited an amateur team and a professional team to participate in the experiments. \hl{When inviting to participate in the experiments, we informed the participants about the data collection process and how their data will be utilized.} Each team consists of five players characterised by similar skill level. The information about participants and their experience is shown in Table~\ref{players_statistics}. The process of the experiment is shown in Figure~\ref{exp}. Game client versions varied from 9.22.296.5720 to 9.24.300.6382, depending on an experimental day.

\begin{table*}[!t]
\centering
\caption{Players' statistics.}
\label{players_statistics}
\begin{tabular}{lllllll}
\multicolumn{1}{l}{Team} & Player id & Hours played & Best rank achieved & Current Rank & Rank percentile & Dominant hand \\\hline\hline
\multirow{5}{*}{Amateurs} & 0 & 336 & Gold & Gold & 34\% & Right \\
 & 1 & 1200 & Gold & Gold & 34\% & Right \\
 & 2 & 1000 & Platinum & Platinum & 9.66\% & Right \\
 & 3 & 400 & No rank & No rank & 100\% & Right \\
 & 4 & 415 & Gold & Gold & 34\% & Right \\\hline
\multirow{5}{*}{Professionals} & 0 & 5000 & Diamond & Diamond & 2.07\% & Left \\
 & 1 & 6000 & Diamond & Diamond & 2.07\% & Right \\
 & 2 & 5000 & Diamond & Diamond & 2.07\% & Right \\
 & 3 & 10000 & Master & Master & 0.09\% & Right \\
 & 4 & 7000 & Diamond & Diamond & 2.07\% & Right\\\hline
 \hline
\end{tabular}
\end{table*}

Each team played up to 4 matches per day on 3 different days. Each match was played either versus bots or real opponents from the Internet; either with or without the team communication (2x2=4 options in total).
On each experimental day, a team played \hl{up to} 4 modifications in random order.
\hl{In total, each team played:}

\begin{itemize}
    \item \hl{3 matches vs real opponents with communication.}
    \item \hl{3 matches vs real opponents without communication.}
    \item \hl{3 matches vs bots with communication.}
    \item \hl{2 matches vs bots without communication (not 3 because of a time limitation in day 2).}
\end{itemize}

\begin{table}[]
\centering
\caption{\hl{Match outcomes for teams.}}
\begin{subtable}{\linewidth}
\centering
\begin{tabular}{c|c|c}
 & Bots & Real Opponents \\ \hline
With communication    & 3/3  & 0/3               \\ \hline
Without communication & 2/2  & 1/3             \\
\end{tabular}
\subcaption{\hl{Match outcomes for the amateur team.}}
\label{match_outcomes_amateurs}
\end{subtable}
\\
\begin{subtable}{\linewidth}
\centering
\begin{tabular}{c|c|c}
 & Bots & Real Opponents \\ \hline
With communication    & 3/3  & 1/3               \\ \hline
Without communication & 2/2  & 2/3             \\
\end{tabular}
\subcaption{\hl{Match outcomes for the professional team.}}
\label{match_outcomes_pros}
\end{subtable}
\end{table}

\hl{Match outcomes for both teams are detailed in Tables}~\ref{match_outcomes_amateurs} and \ref{match_outcomes_pros}.

In bot games, the difficulty of bots was selected to ''Intermediate,`` which was the highest level offered by League of Legends at the time of the experiment.
Games versus bots help to establish the baseline to properly compare the amateur and the professional team. Participants won all matches versus bots. In games versus real opponents, a team played against five people from the Internet 
selected by an in-game matchmaking algorithm.
The matchmaking algorithm tries to find opponents similar to participants in terms of skill.

The reason to record games with or without communication is to check the importance of cooperation for teams of different skill. According to the prior research, quality of team communication might significantly impact eSports team performance \cite{esports_communication}.


After each match had finished, each player passed a short survey and indicated mental load in the game, the level of disturbance and obtrusiveness of the sensing system, and also estimated a personal performance and the performance of teammates.

According to these reports, in 17.5\% of matches, players were not concerned about the sensing system at all; in 69.7\% of matches, they indicated a little bit of disturbance; in 12.8\% of matches, they indicated a significant discomfort caused by the sensing system.

\section{Data Collected}

\begin{figure}[!t]
	\centerline{\includegraphics[width=0.46\linewidth]{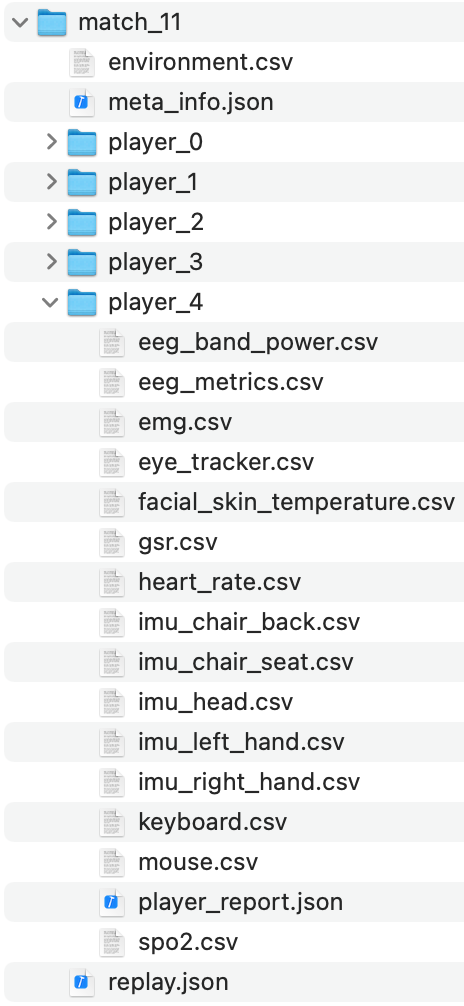}}
	\caption{Dataset structure.}
	\label{dataset_structure}
\end{figure}

\subsection{Dataset Structure}

The dataset consists of data collected in 22 matches and anonymized participants' data. Data about each match are composed of information obtained from in-game logs, sensors, and surveys. The dataset structure is shown in Fig.~\ref{dataset_structure}.
The dataset consists of 22 directories for each match. Match directories include

\begin{itemize}
\item Meta-information file \texttt{meta\_info.json} specifying a team (amateurs or professionals), type of opponents (bots or real players), whether team communication were allowed,
a match outcome, match duration, etc.

\item Match replay in \texttt{replay.json} file with information about game events, such as kill, deaths, items, abilities learned, etc.

\item Five directories with sensor data and post-match surveys for each player. Directories are named as \texttt{player\_\textbf{ID}}, where \texttt{\textbf{ID}} is a player id from 0 to 4.

\item A file with environmental data \texttt{environment.csv}.

\end{itemize}

Each directory \texttt{player\_\textbf{ID}} with data collected from a player includes:

\begin{itemize}

\item A number of \texttt{.csv} files with sensor data.

\item A \texttt{player\_report.json} file with
players' reports on the post-match survey.

\end{itemize}

\begin{table*}[!tp]
\centering
\caption{Collected Sensor Data.}
\label{sensor_data_collected}
\begin{tabular}{lp{10cm}p{2.6cm}}
Sensor Data & Description & Corresponding Files \\\hline\hline
Hand movements &
Data about hand movements were obtained by two IMUs located on both hands
and
included linear acceleration, gravity, angular velocity, Euler angles, and quaternions. Axes orientation for IMUs is shown in Fig.~\ref{imu_wrist}. &
imu\_left\_hand.csv, imu\_right\_hand.csv
\\\hline
Chair movements &
Chair movements were captured by two IMUs attached to the bottom of the chair seat and the chair back. Data logged include linear acceleration, gravity, angular velocity, Euler angles, and quaternions. Axes orientation for IMUs is shown in Fig.~\ref{imu_chair}. &
imu\_chair\_seat.csv, imu\_chair\_back.csv
\\\hline
Head movements &
These data are collected by IMU integrated into the EEG headset. Data recorded include linear acceleration, magnetometer data, and quaternions. &
imu\_head.csv
\\\hline
Gaze &
Eyetracker captured the gaze position, pupil diameter, as well as the validity of these data. &
eye\_tracker.csv
\\\hline
Electrodermal activity &
Data about the electrodermal activity are presented in terms of resistance measured in Ohms. &
gsr.csv
\\\hline
Muscle activity &
Muscle activity is measured by EMG sensor and represented as voltage. &
emg.csv
\\\hline
Heart rate &
Pulse data were collected by the heart rate monitor on the arm and measured in beats per minute \hl{(not the waveform of the pulse oximetry signal)}. &
heart\_rate.csv
\\\hline
EEG &
The signal is
captured by EEG headsets.
Data collected include \hl{alpha, beta, gamma, and theta} waves connected with different types of brain activity. 
Headsets also provide metrics obtained from EEG data (e.g. Engagement, Stress, Focus).
&
eeg\_band\_power.csv,
eeg\_metrics.csv
\\\hline
Face temperature &
Face temperature is presented as the 95-th percentile of values in
48x64 arrays obtained by a thermal camera directed to a player's face. Resulting temperature correspond to player's facial skin temperature.&
face\_temperature.csv
\\\hline
Keyboard activity &
Data about keyboard activity are presented as the number of buttons pressed in the last 5 seconds.
&
keyboard.csv
\\\hline
Mouse activity &
Data about mouse activity are presented as the number of clicks and the distance passed in the last 5 seconds.
&
mouse.csv
\\\hline
Oxygen saturation &
Oxygen saturation is calculated based on reflections of red and infrared light captured by a particle sensor located on the right earlobe.
&
spo2.csv
\\\hline
Environmental data &
Data about temperature, pressure, altitude, humidity, and CO$_2$ level.
All players share the same recordings of environmental data.
&
environment.csv
\\\hline\hline
\end{tabular}
\end{table*}

\subsection{In-game Data}

We obtained in-game logs using the Riot API\footnote{\url{https://developer.riotgames.com/}}. The data are match timelines, including information about key game events and statistics. 
Records collected include the following events:

\begin{enumerate}
    \item \textit{CHAMPION\_KILL}. This event indicates the killer, the victim, the assistants, as well as the location and the timestamp.
    \item \textit{BUILDING\_KILL}. The record specifies the type and the location of the building destroyed, players participated, and the timestamp.
    \item \textit{SKILL\_LEVEL\_UP}. The entry indicates which champion's skill player chooses to upgrade.
    \item \textit{WARD\_PLACED, WARD\_KILL}. These events specify the ward type, the timestamp, and the player involved.
    \item \textit{ITEM\_PURCHASED, ITEM\_DESTROYED}. The records indicate the item, the timestamp, and the player involved.
\end{enumerate}

In-game data are available in the \texttt{replay.json} file.

\subsection{Sensor Data}

\begin{figure}[!t]
	\centering
	\begin{subfigure}[!tp]{0.22\textwidth}
	\centerline{\includegraphics[height=3.5cm]{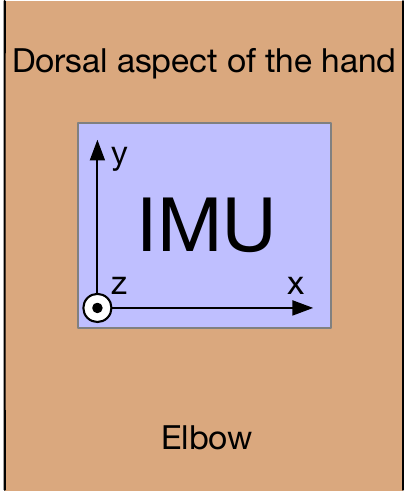}}
	\caption{Orientation of IMU on\\ a wrist.}
	\label{imu_wrist}
	\end{subfigure}
	\begin{subfigure}[!tp]{0.22\textwidth}
	\centerline{\includegraphics[height=3.5cm]{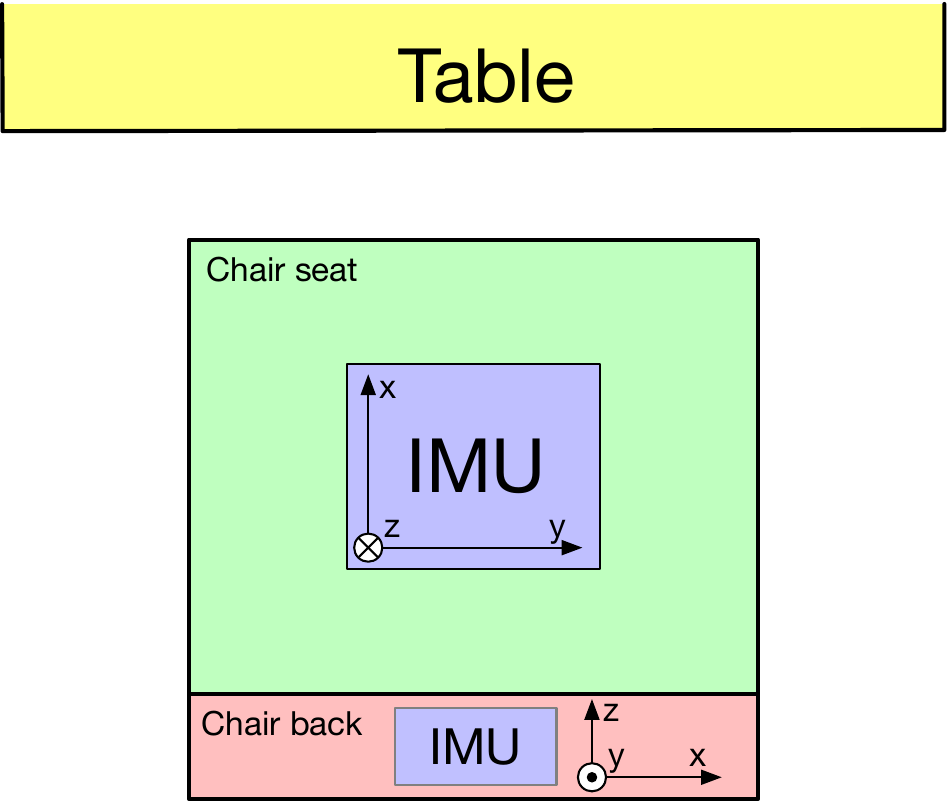}}
	\caption{Orientation of IMUs on a~chair.}
	\label{imu_chair}
    \end{subfigure}
	\caption{Axes orientation for IMUs located on wrists, chair seat, and chair back.}
	\label{sensing_system}
\end{figure}

\begin{figure*}[!b]
	\centerline{\includegraphics[width=\linewidth]{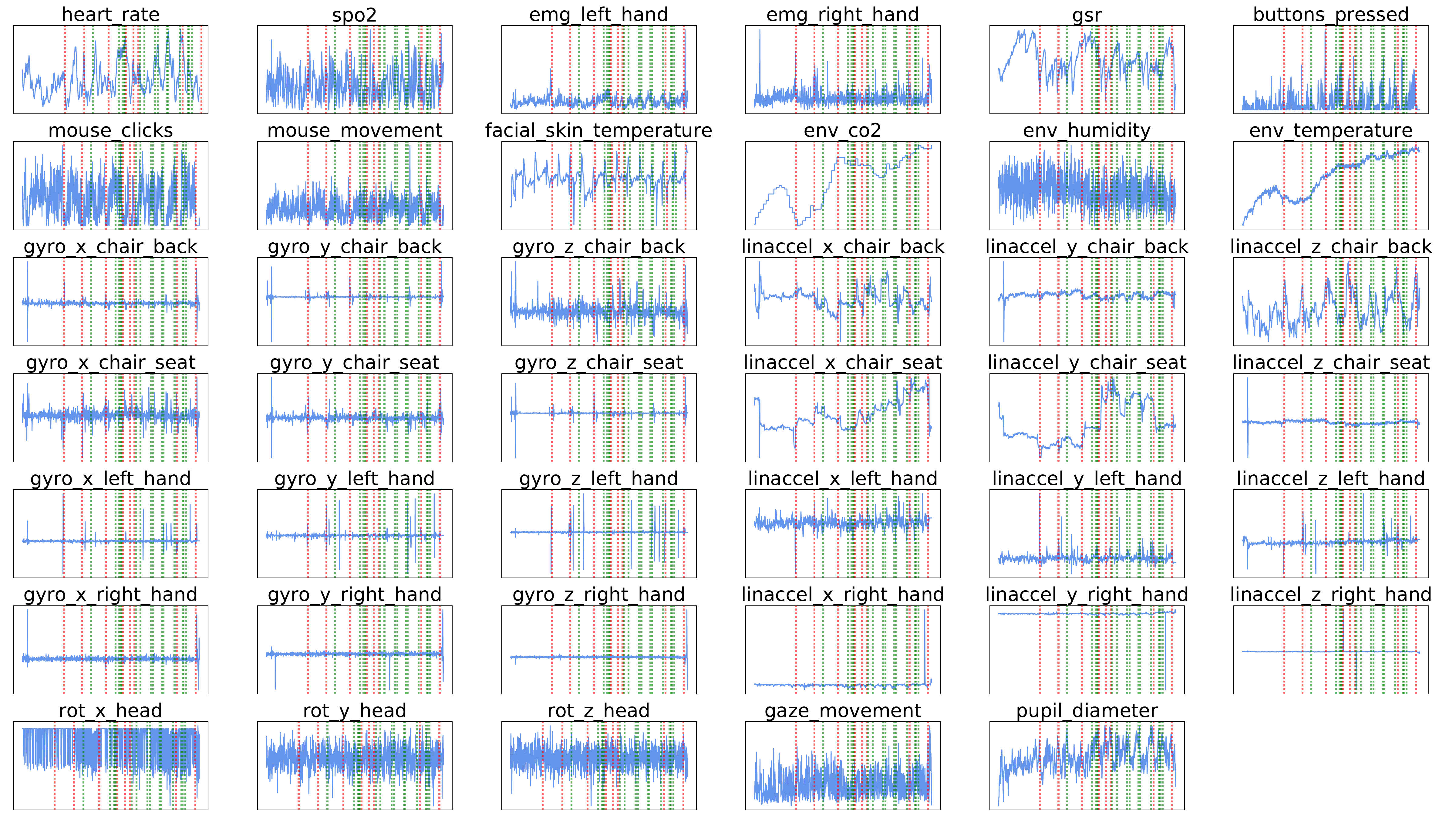}}
	\caption{Selected sensor data collected for one player in one match. Green vertical lines correspond to \texttt{kill}/\texttt{assist} events, red lines correspond to \texttt{death} events.}
	\label{sensor_data_visualization}
\end{figure*}

\begin{figure}[!b]
	\centerline{\includegraphics[width=0.9\linewidth]{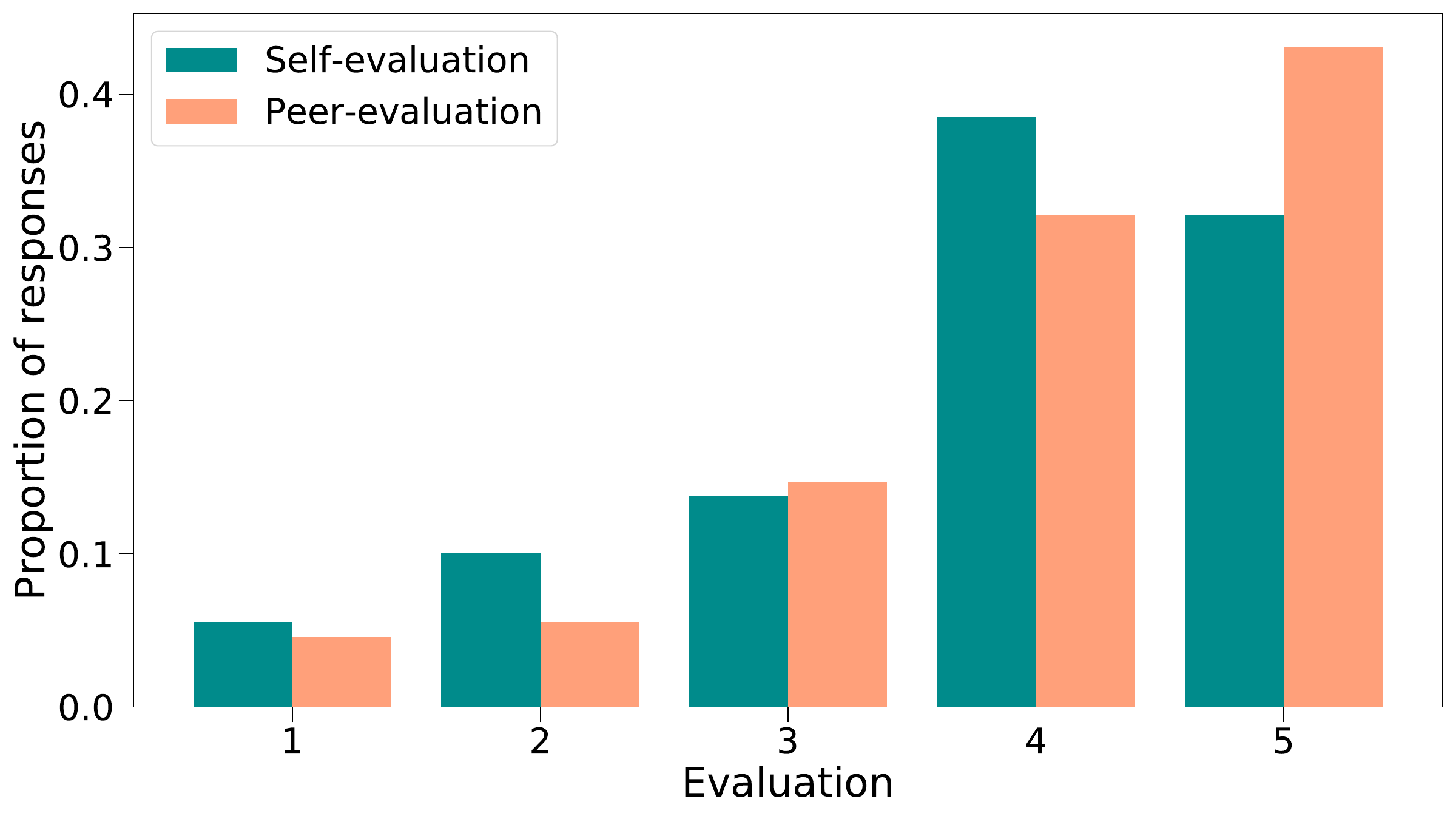}}
	\caption{Distributions of players' responses on teammates performance and self-performance.}
	\label{survey_evaluation}
\end{figure}

\begin{figure}[!b]
	\centerline{\includegraphics[width=0.9\linewidth]{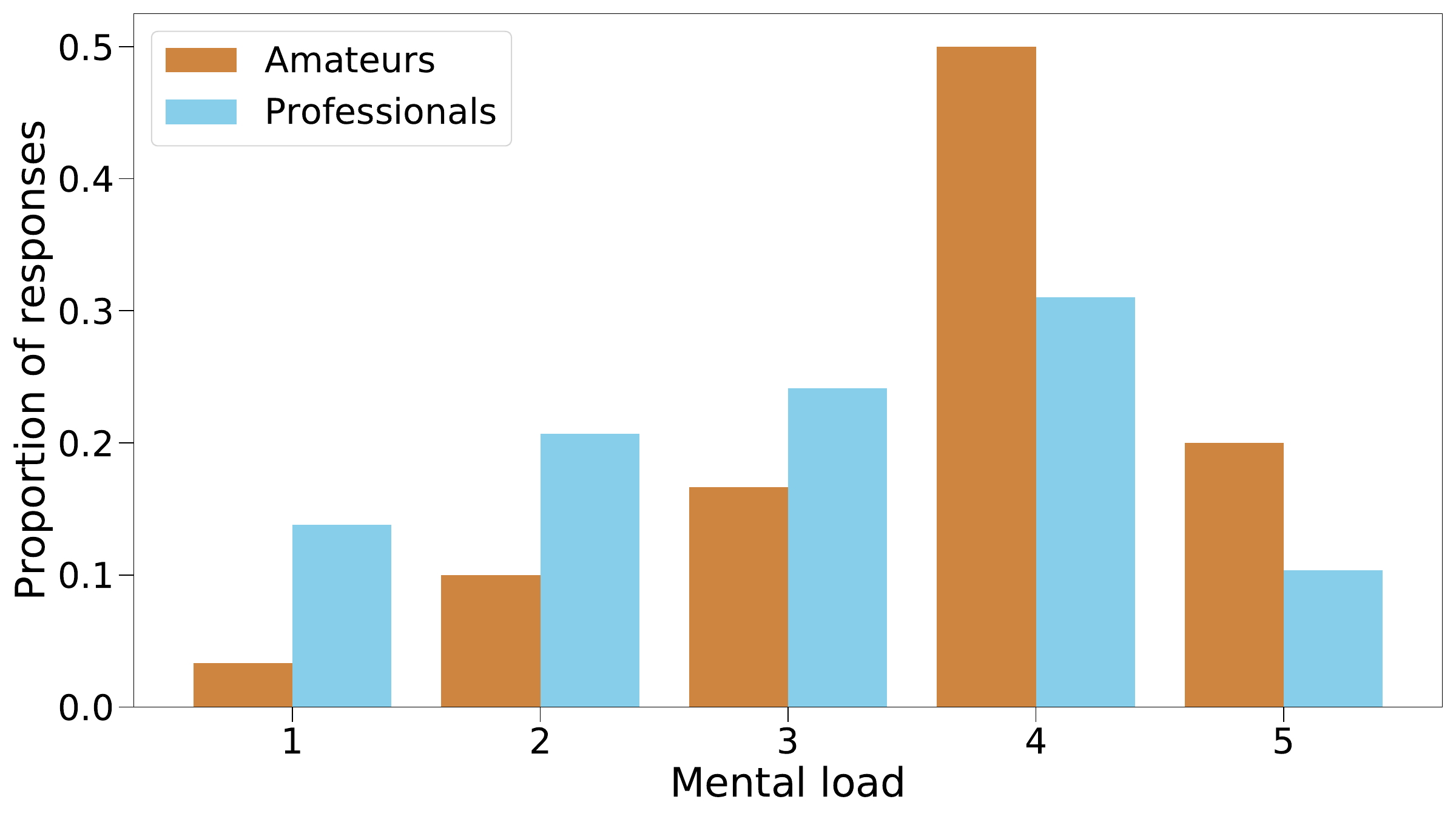}}
	\caption{Distributions of responses on mental load for amateur and professional teams in matches against real opponents.}
	\label{mental_load}
\end{figure}

The overview of collected sensor data is shown in Table~\ref{sensor_data_collected}.
This table describes which signal is measured and how corresponding files are named in the dataset. In addition to the actual sensor signal, each record includes a relative timestamp of the measurement.

The sensor data collected are presented in Fig.~\ref{sensor_data_visualization}. Visualizations of EEG data collected are presented in Appendix \ref{Appendix}. 

These data were processed to remove the noise and outliers.
 Outliers for each sensor were removed by clipping values to an interval between 0.5 and 99.5 percentiles.  
Then the signal was smoothed using an exponential moving average with a half-live value of 1 second. Finally, all signals were resampled to a unified timestep of 1 second by averaging or summation, depending on the nature of source sensor data.

\subsection{Self-reported Survey}

We collected 109 (out of 110 possible) after-match responses from all players, as described in Section~\ref{section_methodology}. In particular, players were asked to evaluate their own performance as well as the performance of their teammates on a scale from 1 to 5. The distribution of results is shown in Figure~\ref{survey_evaluation}. Interestingly, players tend to evaluate the performance of their peers higher than their own performance.

Players also reported mental load in the game on a scale from 1 to 5. The average mental load in games versus bots is only 1.2. In matches against real players, the average mental load is about 3.39. The distribution of responses about the mental load in matches against real opponents is shown in Figure~\ref{mental_load}. 
Clearly, amateurs' mental load is significantly higher than for professionals, even considering the match-making algorithm paired our participants with similarly skilled players. 

\section{Data analysis}

To show the validity of collected data, we perform the analysis of the dataset. We explore our multi-modal dataset to investigate three eSports research topics regarding the player: skill prediction, player re-identification, and team dynamics.

\hl{We used \texttt{python}} \footnote{\url{https://www.python.org/}}
\hl{package
\texttt{pandas}} \footnote{\url{https://pandas.pydata.org/}}
\hl{to preprocess the data and \texttt{scikit-learn}}
\footnote{\url{https://scikit-learn.org/}}
\hl{for the implementations of machine learning models and evaluation metrics.}

\hl{In the Subsection}~\ref{skill_prediction} \hl{we provide results for binary classification of players (pro/amateur) based on sensor data. In the Subsection}~\ref{reid_section} \hl{we demonstrates the possibility of player-reidentification based on sensor data. Finally, in Subsection}~\ref{team_dynamics_section} \hl{we analyse the differences in team dynamics for the amateur and the professional teams and investigate how lack of communication or playing against real opponents affect it.}

\subsection{Skill Prediction} \label{skill_prediction}

Given the sensor data collected from amateur and professional players, a straightforward question is whether it is possible to distinguish them using only the sensor data. For this purpose, we split data collected for each match into 3-minute sessions, up to 5 sessions per match. In total, we got 540 sessions for both teams. Sessions collected in the first two experimental days were used for training, while sessions from the third experimental day were used for testing.

\begin{figure}[!t]
    \begin{subfigure}[!tp]{\linewidth}
    \centerline{\includegraphics[width=0.95\linewidth]{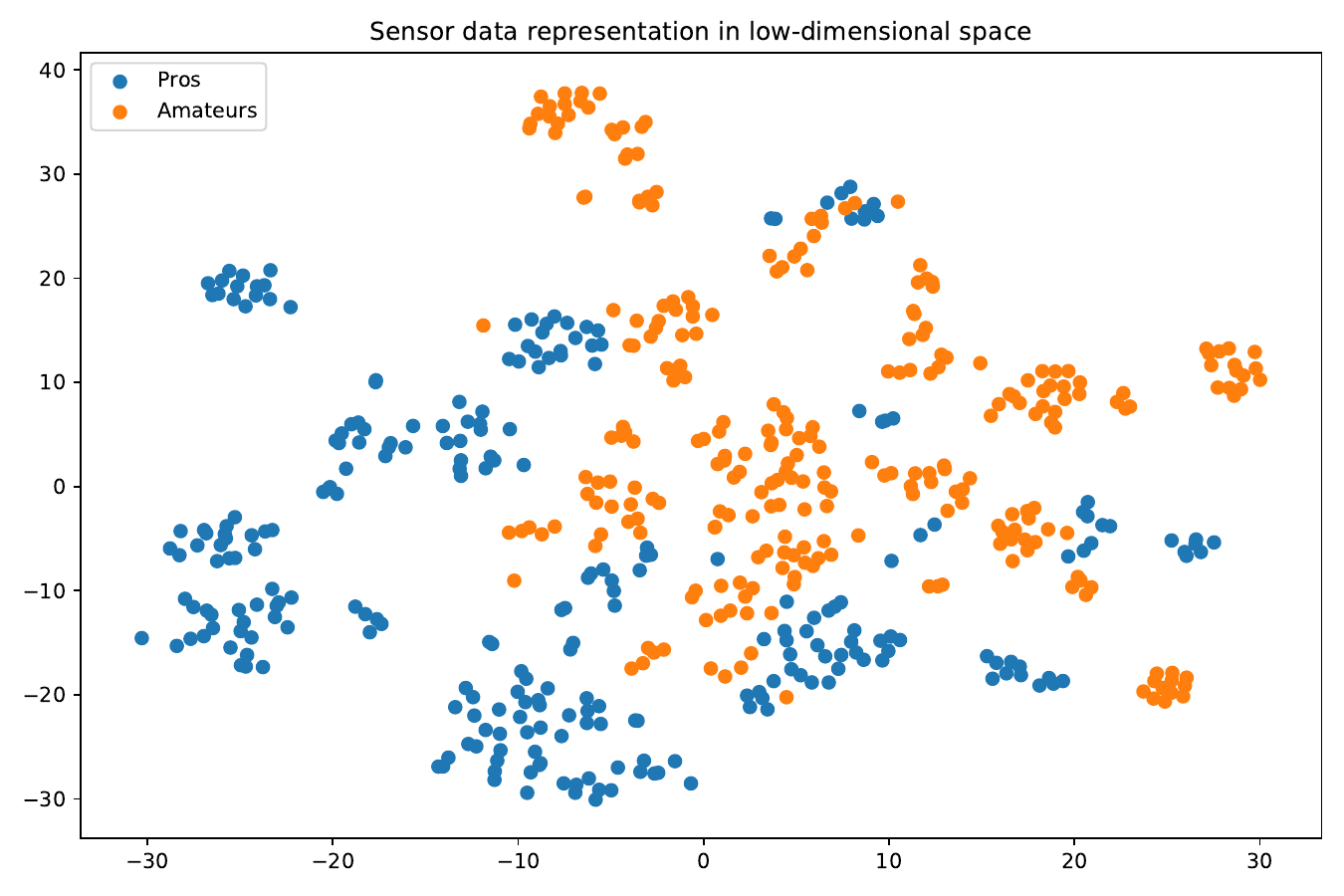}}
	\caption{t-SNE embeddings colored w.r.t. player skill.}
	\label{tsne_skill}
	\end{subfigure}
	\begin{subfigure}[!tp]{\linewidth}
	\centerline{\includegraphics[width=0.95\linewidth]{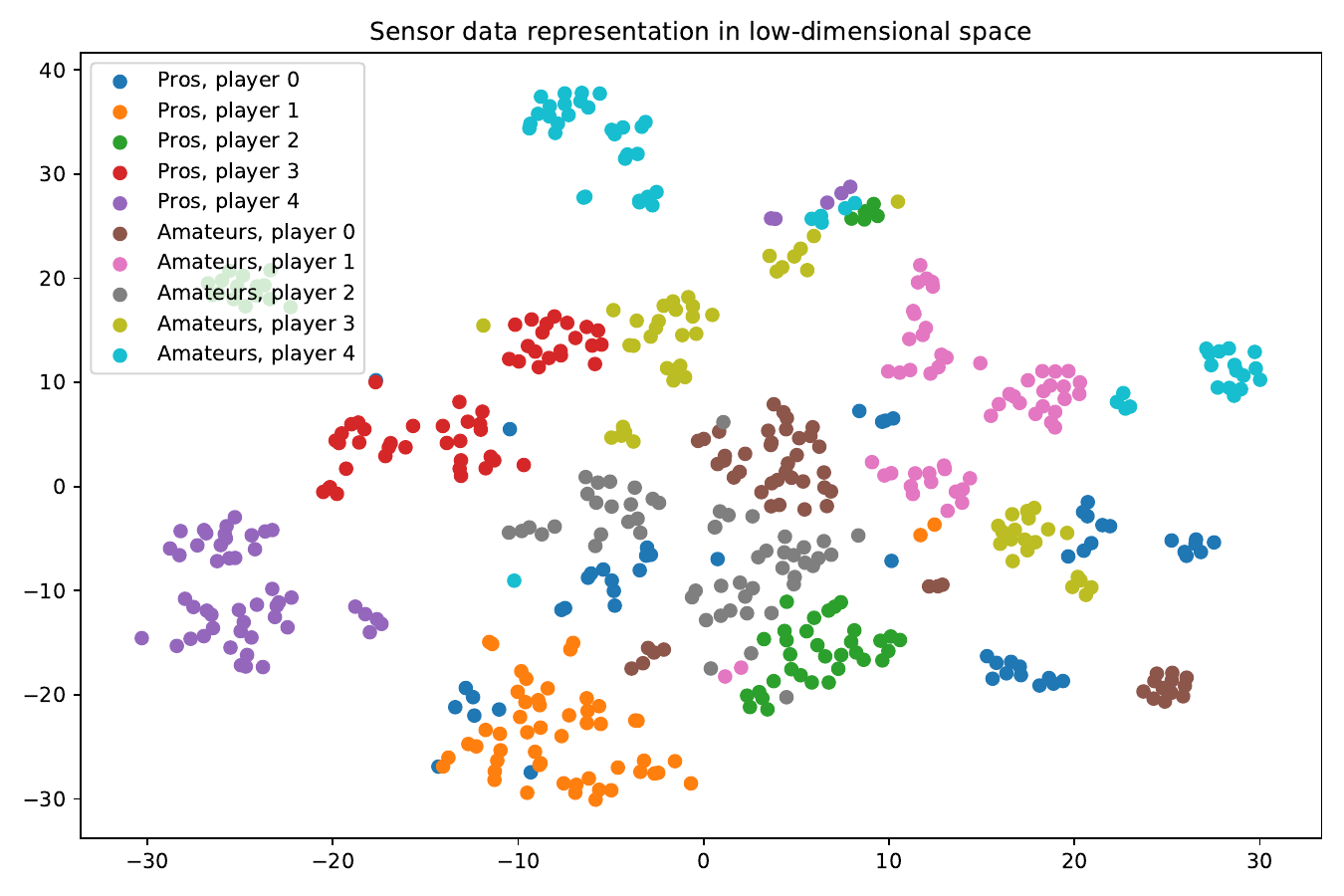}}
	\caption{t-SNE embeddings colored w.r.t. player ID.}
	\label{tsne_reid}
	\end{subfigure}
	\caption{Sensor data embedded in 2-dimensional space.}
	\label{tsne_all}
\end{figure}

To extract features from each session, we averaged sensor signal over time 
and got a 37-dimensional representation for each session (here we didn't use EEG data and environmental data).
For each player, each feature in the representation was normalized to zero mean and unit variance.
We visualized these multidimensional representations obtained in Fig.~\ref{tsne_skill} and Fig.~\ref{tsne_reid} using 2-dimensional embeddings obtained by the t-SNE method \cite{tsne}. Fig.~\ref{tsne_skill} is colored w.r.t. to player skill; Fig.~\ref{tsne_reid} is colored w.r.t. to player ID.

Figure~\ref{tsne_skill} demonstrates that sensor data from players with high and low skill levels reside in different regions in the feature space; thus, it may be separated with a classifier.

We trained several classical machine learning algorithms to predict player skill using 37-dimensional feature vectors obtained from sensor data:

\begin{enumerate}
    \item Logistic regression. A simple and robust linear model for classification \cite{logistic_regression},
    \item k-nearest neighbors classifier \cite{knn} with k=16. This algorithm makes predictions based on similarity in feature space.
    We found k=16 to better suit our problem.
    \item Support Vector Machine with radial basis functions (RBF) kernel. A nonlinear method for robust data classification \cite{svm}.
    \item Random forest classifier \cite{random_forest} with 100 estimators and maximum tree depth 4. An ensemble of diverse decision trees.
    \item Naive Bayes \cite{naive_bayes}. This method makes predictions assuming that features are independent.
    We used Gaussian Naive Bayes since sensor data obtained are continuous. 
    \item Gaussian process \cite{gaussian_process}. 
    This algorithm tries to predict the target class using a latent function. 
\end{enumerate}

For evaluation, we used the following metrics:

\begin{enumerate}
    \item Accuracy score \cite{accuracy}. It is calculated as a proportion of right predictions. The higher values are, the better.
    \item The area under the receiver operating characteristic curve (ROC AUC) \cite{roc_auc}. It takes values from 0 to 1 with a 0.5 score for random guessing. The higher values are, the better.
    \item Log Loss, or binary cross-entropy \cite{logloss}. This loss function is widely utilized in information theory. The lower values are, the better.
\end{enumerate}

Evaluation results are shown in Table~\ref{results_skill}. We used accuracy, logloss, and ROC AUC metrics for evaluation. The target is balanced, so these metrics are suitable for the problem.

The best evaluation quality is achieved by the SVM method: 0.856 accuracy, 0.945 ROC AUC, and 0.311 logloss scores.
In other words, that is possible to estimate player skill by sensor data if the player was presented in a training set.
This model can help investigate how player skill changes over time to provide quick feedback and analytics.

\begin{table}[!t]
\centering
\caption{Evaluation of machine learning methods for skill prediction. Scores for a random guess are added for comparison.}
\begin{tabular}{l|c|c|c}
Method                 & Accuracy       & ROC AUC            & Logloss        \\ \hline
Logistic regression    & 0.838          & 0.886          & 0.596          \\ \hline
k-nearest neighbors    & 0.741          & 0.899          & 0.442          \\ \hline
Support vector machine & \textbf{0.856} & \textbf{0.945} & \textbf{0.311} \\ \hline
Random forest          & 0.800          & 0.885          & 0.456          \\ \hline
Naive Bayes            & 0.706          & 0.780          & 4.627          \\ \hline
Gaussian process       & 0.791          & 0.835          & 0.693
    \\ \hline
Random guess       & 0.500          & 0.500          & 0.693
\label{results_skill}
\end{tabular}
\end{table}

\subsection{Player Re-Identification} \label{reid_section}

Another potential application of sensor data is player re-identification. Re-id methods try to identify a player from a database using a data footprint. Such models can be helpful in cheating detection (e.g. when a person plays instead of another player in a tournament), an adaptation of predictive models, or biometric identification. 2-dimensional embeddings obtained by t-SNE (see Fig.~\ref{tsne_reid}) suggest that players may be separated by a classifier.

As our dataset was acquired in a relatively short time span of several weeks, re-id in this case does not reflect the change of playing style or skill improvement over time. But it is interesting to see how players' skill improvement and changes of playstyles can be reflected from the sensor data.

\begin{table}[!b]
\centering
\caption{Evaluation of machine learning methods for player re-identification. Scores for a random guess are added for comparison.}
\begin{tabular}{l|l|l|l}
Method                 & Accuracy       & ROC AUC            & Logloss        \\ \hline
Logistic regression    & 0.488          & 0.884          & 1.615          \\ \hline
k-nearest neighbors    & 0.415          & 0.840          & 5.735          \\ \hline
Support vector machine & 0.450          & 0.890          & \textbf{1.588} \\ \hline
Random forest          & \textbf{0.521} & \textbf{0.919} & 1.617          \\ \hline
Naive Bayes            & 0.341          & 0.686          & 22.239         \\ \hline
Gaussian process       & 0.447          & 0.812          & 2.302          \\ \hline
Random guess           & 0.100            & 0.500            & 2.303
\end{tabular}
\label{results_reid}
\end{table}




We trained machine learning models on the sensor data described in Section~\ref{skill_prediction} using player id as a target. Models predicted one of 10 classes corresponding to players. The results are shown in Table~\ref{results_reid}. We used the One-vs-Rest strategy to calculate ROC AUC for multiclass classification \cite{auc_multiclass}.

The scores presented show that it is possible to identify a player with 0.521 accuracy, 0.919 ROC AUC, and 1.588 logloss. Relatively high scores for accuracy and ROC AUC demonstrate the possibility of identifying the actual player in the top several predictions.

\subsection{Team Dynamics}\label{team_dynamics_section}

A straightforward approach to measure team dynamics in terms of sensor data is to calculate pairwise correlations \cite{team_dynamics_correlation} to estimate how synchronized players physiology. 
We calculated team dynamics as \hl{average} pairwise \hl{Pearson} correlations for different features extracted from sensor data.
\hl{That means that for each feature
we calculated the correlation for each pair of players and then averaged these values for all pairs. The result represents how each feature is synchronized for the whole team. We used \texttt{numpy.corrcoef} \footnote{https://numpy.org/doc/stable/reference/generated/numpy.corrcoef.html} function to calculate each pairwise correlation. We also estimated errors by checking how results change when calculated on random subsets containing 80\% of matches. For all features reported the error doesn't exceed 0.03.}

\begin{figure}[!b]
	\centerline{\includegraphics[width=\linewidth]{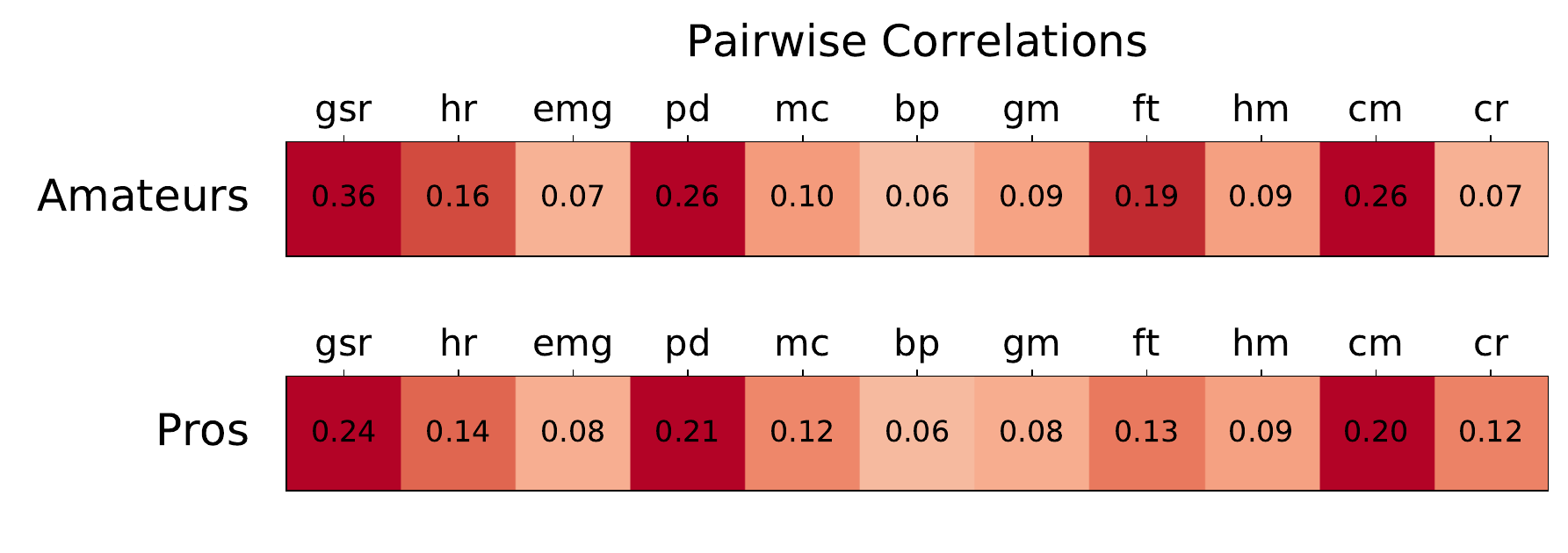}}
	\caption{Team dynamics as \hl{average} pairwise correlations for sensor data \hl{for all matches played}.}
	\label{team_dynamics}
\end{figure}

Results are presented in Fig.~\ref{team_dynamics}; abbreviations are explained in Table~\ref{abbreviations}.

\begin{table}[!b]
\centering
\caption{Feature Abbreviations.}
\begin{tabular}{c|l|p{3.4cm}}
Abbreviation & Description           & Possible Interpretation \\\hline
gsr          & Galvanic skin response                      & A measure of stress/calmness.                        \\\hline
hr           & Heart rate            &                        A measure of stress/calmness. \\\hline
\multirow{3}{*}{emg}          & \multirow{3}{*}{Electromyography}      &                      The intensity of the player's muscle activity (here we used data from the right hand). \\\hline
\multirow{3}{*}{pd}           & \multirow{3}{*}{Pupil diameter}        & A measure of player's concentration and cognitive load \cite{pupil_dilation_concentration, pupil_size_mental_load}.      \\\hline
mp           & Mouse clicks         &                        Frequency of mouse clicks.  \\\hline
\multirow{2}{*}{bp}           & \multirow{2}{*}{Buttons pressed}        & Frequency of keyboard strokes. \\\hline
gm           & Gaze movement         & Average saccade speed.  \\\hline
\multirow{2}{*}{ft}           & \multirow{2}{*}{Face temperature}      & A measure of mental load \cite{face_temperature_mental_load}    \\\hline
\multirow{3}{*}{hm}           &  \multirow{3}{*}{Hand linear movement}  & The intensity of hand movement (here we used data from the right hand). \\\hline
\multirow{2}{*}{cm}           & \multirow{2}{*}{Chair linear movement} &  How intensely the player moves in a chair. \\\hline
\multirow{2}{*}{cr}           & \multirow{2}{*}{Chair rotation}        &  How intensely the player spins on a chair.                      
\end{tabular}
\label{abbreviations}
\end{table}

In general, physiology for the amateur team is more correlated than for the professional team. That particularly holds for galvanic skin response signal, facial skin temperature, and chair movements. Lower correlations for professionals might imply a more independent playstyle of each player, which is not affected by teammates' mistakes or successes.

All the features are positively-correlated For both teams, which is self-explanatory, because players are viewing and participating in the same match. However, some features like pupil diameter, galvanic skin response, and chair movements
are more correlated than others,
and some features like EMG, keyboard activity, and gaze movement are less correlated than others. The possible explanation is that players share a similar level of stress and concentration, although the number and frequency of keystrokes, clicks, and saccades don't not correlate much because they are only the medium of controlling the game character.



To investigate the difference between two teams more precisely, we checked how the presence or absence of communication impacts team dynamics.
We also compared how games against real people instead of bots change teams' behavior.

\begin{figure}[!h]
    \begin{subfigure}[!tp]{\linewidth}
	\centerline{\includegraphics[width=\linewidth]{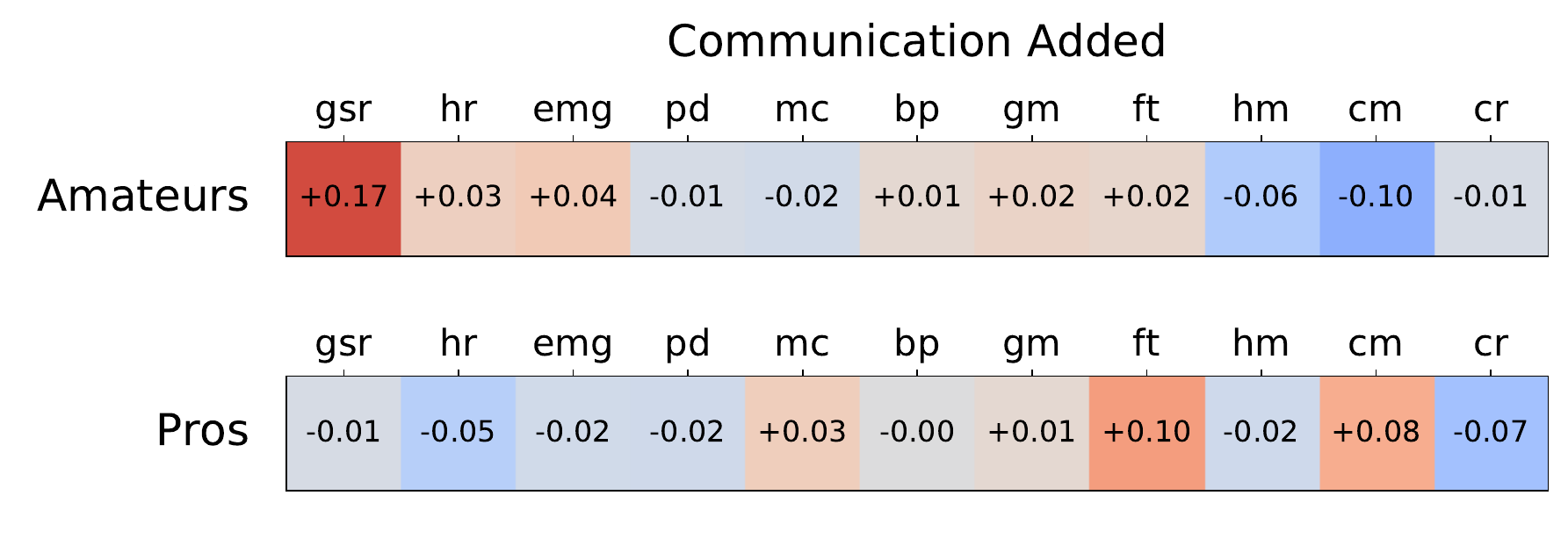}}
	\caption{Changes in \hl{average} pairwise correlations in matches with allowed communication.}
	\label{team_dynamics_comm}
	\end{subfigure}
	\begin{subfigure}[!tp]{\linewidth}
	\centerline{\includegraphics[width=\linewidth]{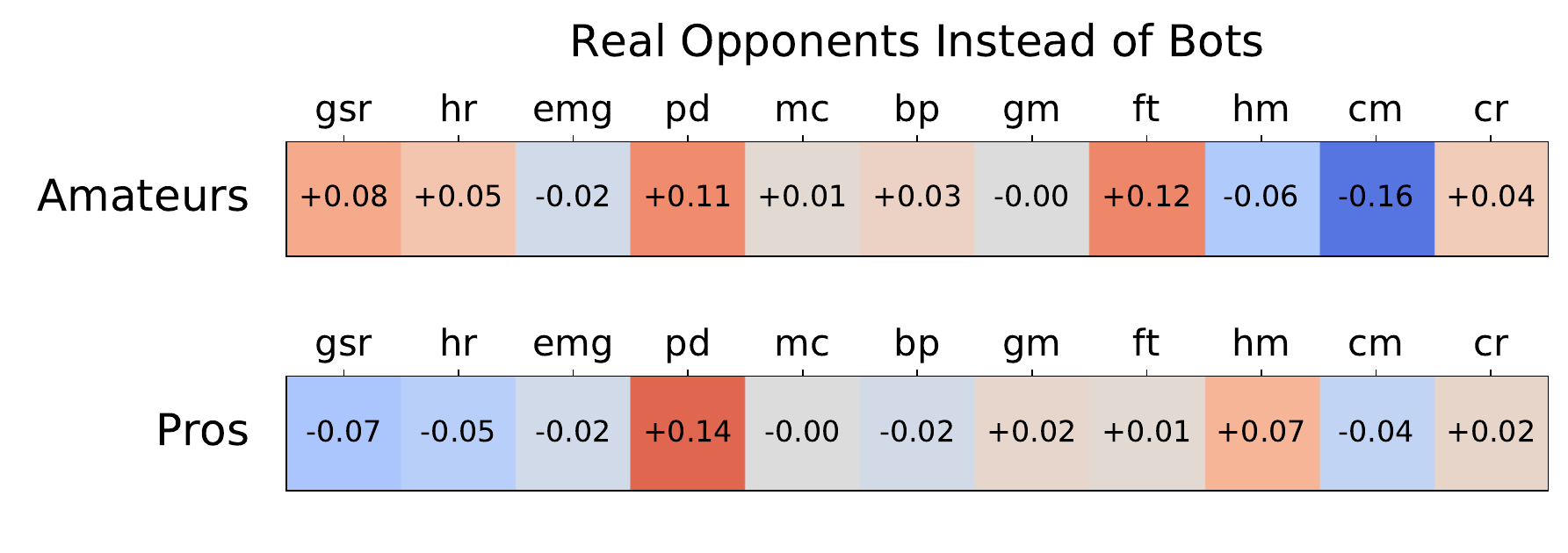}}
	\caption{Changes in \hl{average} pairwise correlations in matches with real opponents.}
	\label{team_dynamics_opp}
	\end{subfigure}
	\caption{Changes in team dynamics in different conditions.}
	\label{team_comp}
\end{figure}

\hl{Fig.}~\ref{team_dynamics_comm} \hl{demonstrates the difference between average pairwise correlations in matches with communication and without it.
Fig.}~\ref{team_dynamics_opp} \hl{shows the difference between matches against real opponents and bots.}


When team communication is allowed, galvanic skin response is significantly more correlated for the amateur team, which might indicate that players share similar stress levels. That is not the case for the professional team, which shows more independent play in terms of stress. 

In matches against real opponents for both teams, the pupil diameter size is more correlated.
Since pupil diameter indicates players' concentration and cognitive load, that implies that players' engagement is more synchronized in matches versus real players rather than versus bots. 
Facial skin temperature data for amateurs are also more correlated, while chair movements are less correlated, which is also showing increased concentration in matches against real players.




\section{Discussion}

\hl{Results in Subsections} \ref{skill_prediction}, \ref{reid_section}, and \ref{team_dynamics_section} \hl{shows the connection between sensor data and in-game outcomes. These might be utilized in many real-world scenarios.}

\hl{Skill prediction system might be utilized for recruiting eSports players, evaluating their skill, or providing feedback on which physiological traits can be improved.
Player re-identification model might help to detect cheaters. An estimation of team dynamics might help to evaluate how players work as a team. All these data might serve as an additional source of information for the coach.

We hope that both players and community will benefit from these changes, because that will helps to bring eSports competition to higher levels.

The results should also hold for other MOBA games (e.g. Dota 2, Heroes of the Storm) because of the similar gameplay, as well as other games of skill (checking this hypothesis in included in out future work). The generalization is achieved by using sensor data instead of data specific to one game. Future work also includes analysing if our findings are applicable in other domains, such as pilot/driver skill estimation, stress monitoring, diseases detection, physical training, etc.}

\hl{The main limitations of the dataset are:}

\begin{itemize}
    \item \hl{Hardware limitations resulting in sometimes incomplete data (e.g. Bluetooth interference for EEG or battery limitations for infrared cameras).}
    \item \hl{Only male participants are presented.}
    \item \hl{Only 10 players participated in the study.}
\end{itemize}

\hl{Future work will focus on these limitations by inviting bigger and diverse set of participants, and using better hardware.}

\section{Conclusions and Future Work}

In this paper, we have presented a dataset with the data from 10 eSports players collected in 22 matches in League of Legends.
Data recorded include players' physiology (e.g. hand, head and chair movements, heart rate, gaze, environmental data), in-game events (e.g. kills, abilities, positioning), self-reported surveys from players (prior experience, evaluation of performance), and other metadata.
The important aspect of the dataset is simultaneous data collection from 5 players, which enables the consequent analysis of team dynamics. 
We have compared an amateur and a professional team and learned that professionals stick to a more independent and homogeneous playstyle.

We have also shown the possibility of utilizing sensor data for skill prediction and player re-identification problems. Best models show a validation set accuracy of 0.856 for player skill prediction and 0.521 for player re-id (0.10 for a random guess).

Given the growing popularity of both eSports and wearable devices,
utilization of sensor data for video games analytics might provide previously inaccessible feedback and insights, and in our work we showed that sensor data help understand the difference between the amateurs and professional players.
Future work includes a further in-depth analysis of the dataset and
investigation of current and new
research directions.
The multi-modal data collected might be used for many applications, such as
encounter outcome prediction, anomaly detection, and performance prediction. We encourage the research community to propose new problems and explore insights based on the dataset collected.

\section*{Acknowledgment}
The reported study was partially funded by RFBR (Russian Foundation for Basic Research) according to the research project No. 18-29-22077.

The research reported in this paper has been partially supported by the BMBF (German Federal Ministry of Education and Research) in the project HeadSense (project number 01IW18001). The support is gratefully acknowledged.

Also, the authors would like to thank the eSports club\footnote{\url{https://www.esports-kl.de/}} at the University of Kaiserslautern for their contribution to methodology development and participation in data collection.

\bibliography{references}{}

\begin{thebibliography}{10}
\providecommand{\url}[1]{#1}
\csname url@samestyle\endcsname
\providecommand{\newblock}{\relax}
\providecommand{\bibinfo}[2]{#2}
\providecommand{\BIBentrySTDinterwordspacing}{\spaceskip=0pt\relax}
\providecommand{\BIBentryALTinterwordstretchfactor}{4}
\providecommand{\BIBentryALTinterwordspacing}{\spaceskip=\fontdimen2\font plus
\BIBentryALTinterwordstretchfactor\fontdimen3\font minus
  \fontdimen4\font\relax}
\providecommand{\BIBforeignlanguage}[2]{{%
\expandafter\ifx\csname l@#1\endcsname\relax
\typeout{** WARNING: IEEEtran.bst: No hyphenation pattern has been}%
\typeout{** loaded for the language `#1'. Using the pattern for}%
\typeout{** the default language instead.}%
\else
\language=\csname l@#1\endcsname
\fi
#2}}
\providecommand{\BIBdecl}{\relax}
\BIBdecl

\bibitem{dota_live_prediction_}
\BIBentryALTinterwordspacing
V.~J. Hodge, S.~M. Devlin, N.~J. Sephton, F.~O. Block, P.~I. Cowling, and
  A.~Drachen, ``Win prediction in multi-player esports: Live professional match
  prediction,'' \emph{IEEE Transactions on Games}, November 2019. [Online].
  Available: \url{http://eprints.whiterose.ac.uk/152931/}
\BIBentrySTDinterwordspacing

\bibitem{dota_draft_prediction_1}
N.~Kinkade, L.~Jolla, and K.~Lim, ``Dota 2 win prediction,'' \emph{Univ Calif},
  vol.~1, pp. 1--13, 2015.

\bibitem{lol_live_prediction_1}
A.~L.~C. Silva, G.~L. Pappa, and L.~Chaimowicz, ``Continuous outcome prediction
  of league of legends competitive matches using recurrent neural networks,''
  in \emph{SBC-Proceedings of SBCGames}, 2018, pp. 2179--2259.

\bibitem{eye_tracking_patterns_2}
G.~Choi and M.~Kim, ``Eye gaze information and game level design according to
  fps gameplay beats,'' \emph{Journal of information and communication
  convergence engineering}, vol.~16, pp. 189--196, 2018.

\bibitem{esports_skill_prediction}
N.~Khromov, A.~Korotin, A.~Lange, A.~Stepanov, E.~Burnaev, and A.~Somov,
  ``Esports athletes and players: a comparative study,'' \emph{IEEE Pervasive
  Computing}, vol.~18, no.~3, pp. 31--39, 2019.

\bibitem{esports_gsr}
P.~M. {Blom}, S.~{Bakkes}, and P.~{Spronck}, ``Towards multi-modal stress
  response modelling in competitive league of legends,'' in \emph{2019 IEEE
  Conference on Games (CoG)}, 2019, pp. 1--4.

\bibitem{gaze_reactions}
D.~{Koposov}, M.~{Semenova}, A.~{Somov}, A.~{Lange}, A.~{Stepanov}, and
  E.~{Burnaev}, ``Analysis of the reaction time of esports players through the
  gaze tracking and personality trait,'' in \emph{2020 IEEE 29th International
  Symposium on Industrial Electronics (ISIE)}, 2020, pp. 1560--1565.

\bibitem{smart_chair_iop}
A.~{Smerdov}, E.~{Burnaev}, and A.~{Somov}, ``esports pro-players behavior
  during the game events: Statistical analysis of data obtained using the smart
  chair,'' in \emph{2019 IEEE SmartWorld, Ubiquitous Intelligence Computing,
  Advanced Trusted Computing, Scalable Computing Communications, Cloud Big Data
  Computing, Internet of People and Smart City Innovation
  (SmartWorld/SCALCOM/UIC/ATC/CBDCom/IOP/SCI)}, 2019, pp. 1768--1775.

\bibitem{esports_eeg_0}
Z.~Minchev, G.~Dukov, and S.~Georgiev, ``Eeg spectral analysis in serious
  gaming: An ad hoc experimental application,'' \emph{Bio Automation}, vol.~13,
  no.~4, pp. 79--88, 2009.

\bibitem{identification_by_mouse}
R.~Kaminsky, M.~Enev, and E.~Andersen, ``Identifying game players with mouse
  biometrics,'' \emph{University of Washington. Technical Report}, 2008.

\bibitem{esports_visual_fixations}
B.~B. Velichkovsky, N.~Khromov, A.~Korotin, E.~Burnaev, and A.~Somov, ``Visual
  fixations duration as an indicator of skill level in esports,'' in \emph{IFIP
  Conference on Human-Computer Interaction}.\hskip 1em plus 0.5em minus
  0.4em\relax Springer, 2019, pp. 397--405.

\bibitem{eye_tracking_patterns_1}
G.~Choi and M.~Kim, ``Eye-movement pattern by playing experience in combat
  system of fps game,'' pp. 52--56, 2015.

\bibitem{esports_eeg_1}
B.~{Meneses-Claudio} and A.~{Roman-Gonzalez}, ``Study of the brain waves for
  the differentiation of gamers category between a newbie and a hardcore in the
  game dota 1,'' in \emph{2018 Congreso Argentino de Ciencias de la
  Informática y Desarrollos de Investigación (CACIDI)}, 2018, pp. 1--4.

\bibitem{smart_chair_wf_iot}
A.~{Smerdov}, A.~{Kiskun}, R.~{Shaniiazov}, A.~{Somov}, and E.~{Burnaev},
  ``Understanding cyber athletes behaviour through a smart chair: Cs:go and
  monolith team scenario,'' in \emph{2019 IEEE 5th World Forum on Internet of
  Things (WF-IoT)}, 2019, pp. 973--978.

\bibitem{lol_streams_emotions}
C.~{Ringer}, J.~A. {Walker}, and M.~A. {Nicolaou}, ``Multimodal joint emotion
  and game context recognition in league of legends livestreams,'' in
  \emph{2019 IEEE Conference on Games (CoG)}, 2019, pp. 1--8.

\bibitem{dota_live_prediction_0}
V.~Hodge, S.~Devlin, N.~Sephton, F.~Block, A.~Drachen, and P.~Cowling, ``Win
  prediction in esports: Mixed-rank match prediction in multi-player online
  battle arena games,'' \emph{arXiv preprint arXiv:1711.06498}, 2017.

\bibitem{dota_live_prediction_1}
Y.~Yang, T.~Qin, and Y.-H. Lei, ``Real-time esports match result prediction,''
  \emph{arXiv preprint arXiv:1701.03162}, 2016.

\bibitem{lol_live_prediction_0}
L.~Lin, ``League of legends match outcome prediction.''

\bibitem{fps_prediction_0}
K.~J. Shim, K.-W. Hsu, S.~Damania, C.~DeLong, and J.~Srivastava, ``An
  exploratory study of player and team performance in multiplayer
  first-person-shooter games,'' in \emph{2011 IEEE Third International
  Conference on Privacy, Security, Risk and Trust and 2011 IEEE Third
  International Conference on Social Computing}.\hskip 1em plus 0.5em minus
  0.4em\relax IEEE, 2011, pp. 617--620.

\bibitem{lol_logic_mining}
L.~C. Kho, M.~S.~M. Kasihmuddin, M.~Mansor, S.~Sathasivam \emph{et~al.},
  ``Logic mining in league of legends.'' \emph{Pertanika Journal of Science \&
  Technology}, vol.~28, no.~1, 2020.

\bibitem{lol_prediction_interpretation}
Z.~Yang, Z.~Pan, Y.~Wang, D.~Cai, S.~Shi, S.-L. Huang, and X.~Liu,
  ``Interpretable real-time win prediction for honor of kings, a popular mobile
  moba esport,'' \emph{arXiv preprint arXiv:2008.06313}, 2020.

\bibitem{encounter_prediction}
M.~Schubert, A.~Drachen, and T.~Mahlmann, ``Esports analytics through encounter
  detection,'' in \emph{Proceedings of the MIT Sloan Sports Analytics
  Conference}, vol.~1.\hskip 1em plus 0.5em minus 0.4em\relax MIT Sloan Boston,
  MA, 2016, p. 2016.

\bibitem{confidence_prediction_lol}
D.-H. Kim, C.~Lee, and K.-S. Chung, ``A confidence-calibrated moba game winner
  predictor,'' \emph{arXiv preprint arXiv:2006.15521}, 2020.

\bibitem{dota_draft_prediction_0}
A.~Semenov, P.~Romov, S.~Korolev, D.~Yashkov, and K.~Neklyudov, ``Performance
  of machine learning algorithms in predicting game outcome from drafts in dota
  2,'' in \emph{International Conference on Analysis of Images, Social Networks
  and Texts}.\hskip 1em plus 0.5em minus 0.4em\relax Springer, 2016, pp.
  26--37.

\bibitem{dota_draft_prediction_2}
N.~Wang, L.~Li, L.~Xiao, G.~Yang, and Y.~Zhou, ``Outcome prediction of dota2
  using machine learning methods,'' in \emph{Proceedings of 2018 International
  Conference on Mathematics and Artificial Intelligence}, 2018, pp. 61--67.

\bibitem{lol_draft_prediction_0}
Z.~Chen, T.-H.~D. Nguyen, Y.~Xu, C.~Amato, S.~Cooper, Y.~Sun, and M.~S.
  El-Nasr, ``The art of drafting: a team-oriented hero recommendation system
  for multiplayer online battle arena games,'' in \emph{Proceedings of the 12th
  ACM Conference on Recommender Systems}, 2018, pp. 200--208.

\bibitem{phys_momentum}
A.~White and D.~M. Romano, ``Scalable psychological momentum forecasting in
  esports,'' \emph{arXiv preprint arXiv:2001.11274}, 2020.

\bibitem{history_prediction_1}
Y.~Yang, T.~Qin, and Y.-H. Lei, ``Real-time esports match result prediction,''
  \emph{arXiv preprint arXiv:1701.03162}, 2016.

\bibitem{history_prediction_2}
I.~Makarov, D.~Savostyanov, B.~Litvyakov, and D.~I. Ignatov, ``Predicting
  winning team and probabilistic ratings in “dota 2” and “counter-strike:
  Global offensive” video games,'' in \emph{International Conference on
  Analysis of Images, Social Networks and Texts}.\hskip 1em plus 0.5em minus
  0.4em\relax Springer, 2017, pp. 183--196.

\bibitem{ffa_rating_system}
A.~Dehpanah, M.~F. Ghori, J.~Gemmell, and B.~Mobasher, ``The evaluation of
  rating systems in online free-for-all games,'' \emph{arXiv preprint
  arXiv:2008.06787}, 2020.

\bibitem{esports_clusters}
A.~Drachen, R.~Sifa, C.~Bauckhage, and C.~Thurau, ``Guns, swords and data:
  Clustering of player behavior in computer games in the wild,'' in \emph{2012
  IEEE conference on Computational Intelligence and Games (CIG)}.\hskip 1em
  plus 0.5em minus 0.4em\relax IEEE, 2012, pp. 163--170.

\bibitem{gao}
L.~Gao, J.~Judd, D.~Wong, and J.~Lowder, ``Classifying dota 2 hero characters
  based on play style and performance,'' \emph{Univ. of Utah Course on ML},
  2013.

\bibitem{dl_bot}
J.~Pfau, J.~D. Smeddinck, I.~Bikas, and R.~Malaka, ``Bot or not? user
  perceptions of player substitution with deep player behavior models,'' in
  \emph{Proceedings of the 2020 CHI Conference on Human Factors in Computing
  Systems}, 2020, pp. 1--10.

\bibitem{toxic_behavior}
S.~Adinolf and S.~Turkay, ``Toxic behaviors in esports games: player
  perceptions and coping strategies,'' in \emph{Proceedings of the 2018 Annual
  Symposium on Computer-Human Interaction in Play Companion Extended
  Abstracts}, 2018, pp. 365--372.

\bibitem{toxicity_detection}
M.~{Märtens}, S.~{Shen}, A.~{Iosup}, and F.~{Kuipers}, ``Toxicity detection in
  multiplayer online games,'' in \emph{2015 International Workshop on Network
  and Systems Support for Games (NetGames)}, 2015, pp. 1--6.

\bibitem{emg_state}
P.~M. Lehrer, D.~M. Batey, R.~L. Woolfolk, A.~Remde, and T.~Garlick, ``The
  effect of repeated tense-release sequences on emg and self-report of muscle
  tension: An evaluation of jacobsonian and post-jacobsonian assumptions about
  progressive relaxation,'' \emph{Psychophysiology}, vol.~25, no.~5, pp.
  562--569, 1988.

\bibitem{gsr_arousal}
D.~C. Fowles, ``The three arousal model: Implications of gray's two-factor
  learning theory for heart rate, electrodermal activity, and psychopathy,''
  \emph{Psychophysiology}, vol.~17, no.~2, pp. 87--104, 1980.

\bibitem{humidity_performance}
M.~Zhu, W.~Liu, and P.~Wargocki, ``Changes in eeg signals during the cognitive
  activity at varying air temperature and relative humidity,'' \emph{Journal of
  Exposure Science \& Environmental Epidemiology}, vol.~30, no.~2, pp.
  285--298, 2020.

\bibitem{env_temperature_performance}
L.~Lan, P.~Wargocki, D.~P. Wyon, and Z.~Lian, ``Effects of thermal discomfort
  in an office on perceived air quality, sbs symptoms, physiological responses,
  and human performance,'' \emph{Indoor air}, vol.~21, no.~5, pp. 376--390,
  2011.

\bibitem{co2_cognitive}
J.~G. Allen, P.~MacNaughton, U.~Satish, S.~Santanam, J.~Vallarino, and J.~D.
  Spengler, ``Associations of cognitive function scores with carbon dioxide,
  ventilation, and volatile organic compound exposures in office workers: a
  controlled exposure study of green and conventional office environments,''
  \emph{Environmental health perspectives}, vol. 124, no.~6, pp. 805--812,
  2016.

\bibitem{eeg_arousal}
G.~Stenberg, ``Personality and the eeg: Arousal and emotional arousability,''
  \emph{Personality and individual differences}, vol.~13, no.~10, pp.
  1097--1113, 1992.

\bibitem{mouse_skill}
D.~Buckley, K.~Chen, and J.~Knowles, ``Predicting skill from gameplay input to
  a first-person shooter,'' in \emph{2013 IEEE Conference on Computational
  Inteligence in Games (CIG)}.\hskip 1em plus 0.5em minus 0.4em\relax IEEE,
  2013, pp. 1--8.

\bibitem{heart_rate_performance}
J.~Taelman, S.~Vandeput, A.~Spaepen, and S.~Van~Huffel, ``Influence of mental
  stress on heart rate and heart rate variability,'' in \emph{4th European
  conference of the international federation for medical and biological
  engineering}.\hskip 1em plus 0.5em minus 0.4em\relax Springer, 2009, pp.
  1366--1369.

\bibitem{oxygen_saturation_performance}
A.~B. Scholey, M.~C. Moss, and K.~Wesnes, ``Oxygen and cognitive performance:
  the temporal relationship between hyperoxia and enhanced memory,''
  \emph{Psychopharmacology}, vol. 140, no.~1, pp. 123--126, 1998.

\bibitem{esports_communication}
M.~J. Smith, P.~D. Birch, and D.~Bright, ``Identifying stressors and coping
  strategies of elite esports competitors,'' \emph{International Journal of
  Gaming and Computer-Mediated Simulations (IJGCMS)}, vol.~11, no.~2, pp.
  22--39, 2019.

\bibitem{tsne}
L.~v.~d. Maaten and G.~Hinton, ``Visualizing data using t-sne,'' \emph{Journal
  of machine learning research}, vol.~9, no. Nov, pp. 2579--2605, 2008.

\bibitem{logistic_regression}
D.~W. Hosmer~Jr, S.~Lemeshow, and R.~X. Sturdivant, \emph{Applied logistic
  regression}.\hskip 1em plus 0.5em minus 0.4em\relax John Wiley \& Sons, 2013,
  vol. 398.

\bibitem{knn}
P.~Cunningham and S.~J. Delany, ``k-nearest neighbour classifiers--,''
  \emph{arXiv preprint arXiv:2004.04523}, 2020.

\bibitem{svm}
S.-i. Amari and S.~Wu, ``Improving support vector machine classifiers by
  modifying kernel functions,'' \emph{Neural Networks}, vol.~12, no.~6, pp.
  783--789, 1999.

\bibitem{random_forest}
A.~Liaw, M.~Wiener \emph{et~al.}, ``Classification and regression by
  randomforest,'' \emph{R news}, vol.~2, no.~3, pp. 18--22, 2002.

\bibitem{naive_bayes}
I.~Rish \emph{et~al.}, ``An empirical study of the naive bayes classifier,'' in
  \emph{IJCAI 2001 workshop on empirical methods in artificial intelligence},
  vol.~3, no.~22, 2001, pp. 41--46.

\bibitem{gaussian_process}
C.~E. Rasmussen, ``Gaussian processes in machine learning,'' in \emph{Summer
  School on Machine Learning}.\hskip 1em plus 0.5em minus 0.4em\relax Springer,
  2003, pp. 63--71.

\bibitem{accuracy}
M.~Sokolova, N.~Japkowicz, and S.~Szpakowicz, ``Beyond accuracy, f-score and
  roc: a family of discriminant measures for performance evaluation,'' in
  \emph{Australasian joint conference on artificial intelligence}.\hskip 1em
  plus 0.5em minus 0.4em\relax Springer, 2006, pp. 1015--1021.

\bibitem{roc_auc}
J.~Fan, S.~Upadhye, and A.~Worster, ``Understanding receiver operating
  characteristic (roc) curves,'' \emph{Canadian Journal of Emergency Medicine},
  vol.~8, no.~1, pp. 19--20, 2006.

\bibitem{logloss}
V.~Vovk, ``The fundamental nature of the log loss function,'' in \emph{Fields
  of Logic and Computation II}.\hskip 1em plus 0.5em minus 0.4em\relax
  Springer, 2015, pp. 307--318.

\bibitem{auc_multiclass}
J.~Hanley and B.~McNeal, ``A simple generalization of the area under the roc
  curve to multiple class classification problems,'' \emph{Radiology}, vol.
  143, pp. 29--36, 1982.

\bibitem{team_dynamics_correlation}
E.~R. Walker, D.~L. Lang, B.~A. Caruso, and L.~Salas-Hern{\'a}ndez, ``\hl{Role
  of team dynamics in the learning process: a mixed-methods evaluation of a
  modified team-based learning approach in a behavioral research methods
  course},'' \emph{Advances in Health Sciences Education}, pp. 1--17, 2019.

\bibitem{pupil_dilation_concentration}
O.~E. Kang, K.~E. Huffer, and T.~P. Wheatley, ``Pupil dilation dynamics track
  attention to high-level information,'' \emph{PloS one}, vol.~9, no.~8, p.
  e102463, 2014.

\bibitem{pupil_size_mental_load}
B.~Stone, M.~Lee, S.~Dennis, and T.~Nettelbeck, ``Pupil size and mental load,''
  in \emph{1st Adelaide Mental Life Conference}, 2004.

\bibitem{face_temperature_mental_load}
T.~Mizuno, T.~Sakai, S.~Kawazura, H.~Asano, K.~Akehi, S.~Matsuno, K.~Mito,
  Y.~Kume, and N.~Itakura, ``Measuring facial skin temperature changes caused
  by mental work-load with infrared thermography,'' \emph{IEEJ Transactions on
  Electronics, Information and Systems}, vol. 136, no.~11, pp. 1581--1585,
  2016.

\end{thebibliography}
\bibliographystyle{IEEEtran}

\begin{appendices}
\section{EEG Data}\label{Appendix}

\begin{figure}[!h]
	\centerline{\includegraphics[width=\linewidth]{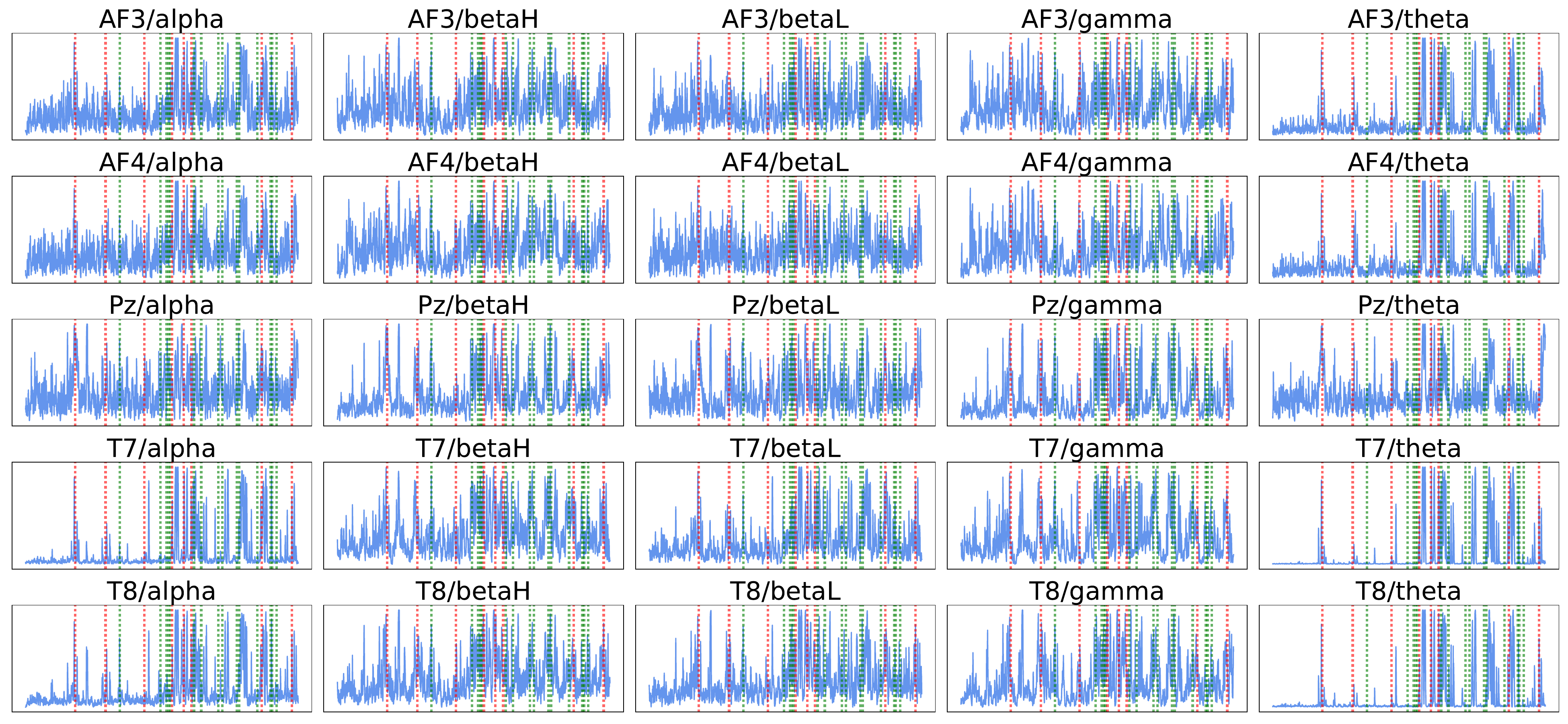}}
	\caption{EEG band power for one match for one player. Green vertical lines correspond to \texttt{kill}/\texttt{assist} events, red lines correspond to \texttt{death} events.}
	\label{eeg_band_power}
\end{figure}

\begin{figure}[!h]
	\centerline{\includegraphics[width=\linewidth]{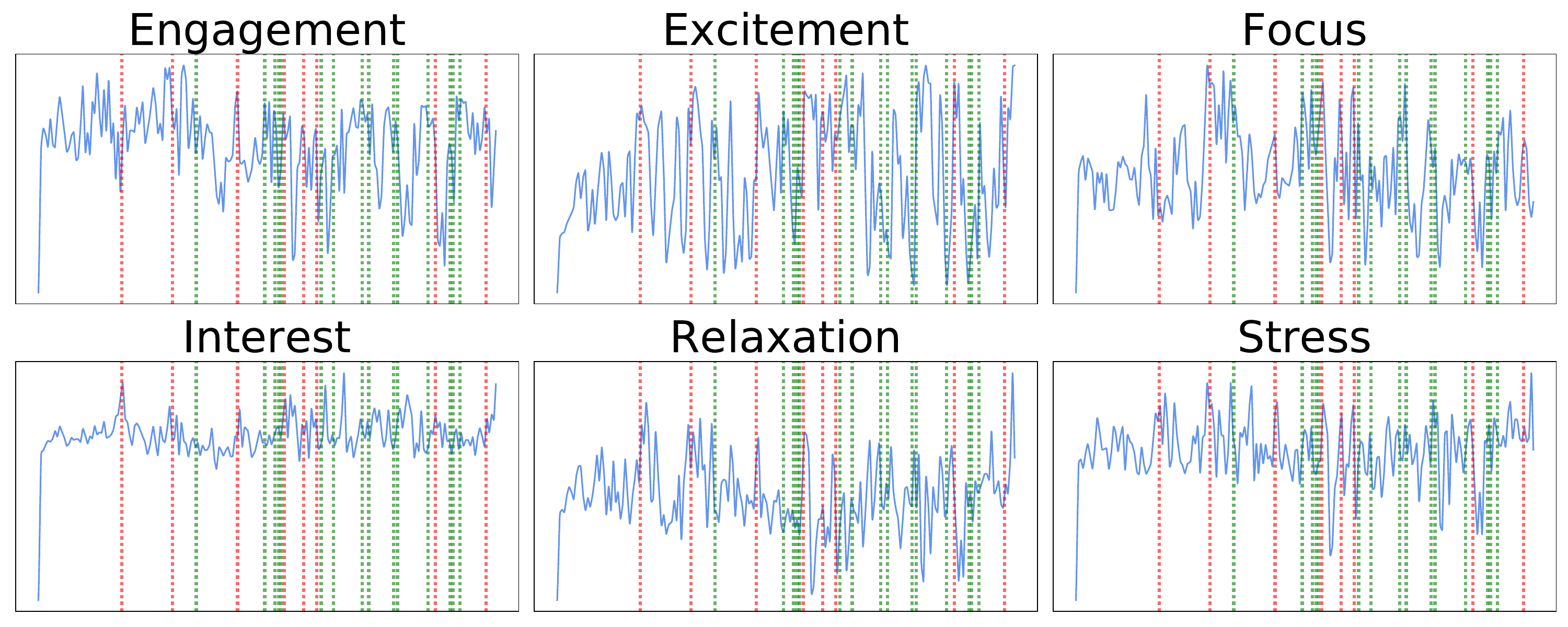}}
	\caption{EEG metrics calculated by the EEG headset for one match for one player. Green vertical lines correspond to \texttt{kill}/\texttt{assist} events, red lines correspond to \texttt{death} events.}
	\label{eeg_metrics}
\end{figure}

\hl{Additional visualizations for EEG band power and EEG performance metrics are shown in} Fig.~\ref{eeg_band_power} and Fig.~\ref{eeg_metrics}.

\hl{EEG band power represents the intensity of include alpha, beta, gamma, and theta waves connected with different types of brain activity. 

EEG performance metrics are calculated by the headset} \footnote{\url{https://www.emotiv.com/knowledge-base/performance-metrics/}} \hl{and provides six basic measures of mental performance. Each measure is automatically scaled to suit participant's normal range and base level of each condition.}

\end{appendices}

\end{document}